\documentclass[conference,compsoc]{IEEEtran}

\usepackage{latexsym,fancyhdr,url}
\usepackage[ruled,vlined,linesnumbered]{algorithm2e}
\usepackage{algorithm2e}
\usepackage{algpseudocode}
\usepackage{booktabs}
\usepackage{xparse} % argument parsing -- \edist
\usepackage{xspace}
\usepackage{multirow}
\usepackage{makecell}
\usepackage{pifont}
\usepackage{enumitem}
\usepackage{xcolor}
\usepackage{graphicx}
\usepackage{hyperref}
\algnewcommand{\algorithmicand}{\textbf{ and }}
\algnewcommand{\algorithmicor}{\textbf{ or }}
\algnewcommand{\OR}{\algorithmicor}
\algnewcommand{\AND}{\algorithmicand}
\algnewcommand{\var}{\texttt}
\definecolor{light-gray}{gray}{0.95}
\usepackage{
  color,
  float,
  epsfig,
  wrapfig,
  graphics,
  graphicx,
  subcaption
}

\usepackage{listings}
\usepackage{xcolor}
\usepackage{caption}
\usepackage{microtype}
\microtypecontext{spacing=false}
\microtypesetup{kerning=false}

\definecolor{codegreen}{rgb}{0,0.6,0}
\definecolor{codegray}{rgb}{0.5,0.5,0.5}
\definecolor{codeblue}{rgb}{0,0,1}
\definecolor{backcolour}{rgb}{0.99,0.99,0.99}

\lstdefinestyle{cpp}{
    backgroundcolor=\color{backcolour},   
    commentstyle=\color{codegray},
    keywordstyle=\color{codeblue},
    numberstyle=\tiny\color{codegray},
    stringstyle=\color{codepurple},
    basicstyle=\ttfamily\footnotesize,    
    breaklines=true,                 
    captionpos=b, 
    numbers=left,
    numbersep=5pt,
}
\makeatletter
\renewcommand\paragraph{\@startsection{paragraph}{4}{\z@}%
  {0.1\baselineskip }%
  {-0.5em}
  {\normalfont\normalsize\bfseries}}
\makeatother

\newcommand{\code}[1]{\lstinline[language=c]{#1}}
\newcommand{\sys}{\texttt{CAMP}\xspace}

\date{}

\title{\Large \bf CAMP: Compiler and Allocator-based Heap Memory Protection}
\author{
{\rm Zhenpeng Lin, Zheng Yu, Ziyi Guo, Simone Campanoni, Peter Dinda, and Xinyu Xing} \\
\textit{\{zplin, zhengyu2027, n7l8m4\}@u.northwestern.edu}\\
\textit{\{simone.campanoni, pdinda, xinyu.xing\}@northwestern.edu}\\
{\rm \textit{Northwestern University}}
}

\usepackage{listings}
\usepackage{caption}

\usepackage{xcolor}
\definecolor{codepurple}{rgb}{0.5,0.0,0.5}

\definecolor{codegreen}{rgb}{0,0.6,0}
\definecolor{codegray}{rgb}{0.5,0.5,0.5}
\definecolor{codeblue}{rgb}{0,0,1}
\definecolor{backcolour}{rgb}{0.99,0.99,0.99}

\lstdefinestyle{cpp}{
    backgroundcolor=\color{backcolour},   
    commentstyle=\color{codegray},
    keywordstyle=\color{codeblue},
    numberstyle=\tiny\color{codegray},
    stringstyle=\color{codepurple},
    basicstyle=\ttfamily\footnotesize,
    breakatwhitespace=false,         
    breaklines=true,                 
    captionpos=b, 
    numbers=left,
    numbersep=5pt,
}
\begin{document}
\maketitle

% Use the following at camera-ready time to suppress page numbers.
% Comment it out when you first submit the paper for review.
\pagestyle{empty}

\subsection*{Abstract}

The heap is a critical and widely used component of many applications. Due to its dynamic nature, combined with the complexity of heap management algorithms, it is also a frequent target for security exploits. To enhance the heap's security, various heap protection techniques have been introduced, but they either introduce significant runtime overhead or have limited protection.  
We present \sys, a new sanitizer for detecting and capturing heap memory corruption. \sys leverages a compiler and a customized memory allocator. The compiler adds boundary-checking and escape-tracking instructions to the target program, while the memory allocator tracks memory ranges, coordinates with the instrumentation, and neutralizes dangling pointers. With the novel error detection scheme, \sys enables various compiler optimization strategies and thus eliminates redundant and unnecessary check instrumentation. This design minimizes runtime overhead without sacrificing security guarantees. Our evaluation and comparison of \sys with existing tools, using both real-world applications and SPEC CPU benchmarks, show that it provides even better heap corruption detection capability with lower runtime overhead.
\section{Introduction}

The heap is a region of memory that is dynamically allocated during runtime. It is widely used for dynamic memory allocation and for storing data structures with variable sizes. Due to its frequent use and complex nature, the heap is vulnerable to spatial and temporal memory errors. The prominence of heap errors as a source of vulnerabilities has been consistently high over the years. According to Microsoft, heap errors were responsible for 53\% of Remote Code Execution (RCE) CVEs in their products~\cite{msr_vuln_stat}. Google Project Zero discovered that heap errors accounted for 69\% of zero-day vulnerabilities observed in the wild~\cite{p0_vuln_stat} in 2022. Until now, 65\% vulnerabilities are confirmed as heap-based zero-day in Linux, 2023~\cite{2023kernelcve}.

Over the years, various heap protection techniques have been introduced to enhance the security of the heap, including improved heap management algorithms~\cite{freeguard} and the implementation of layout randomization techniques~\cite{lin2009polymorphing, dieharder}. These advancements have made the heap more secure. However, the continuous discovery and development of new vulnerabilities and exploitation techniques~\cite{wang2021maze, heelan2018automatic, lin2022dirtycred, yun2020automatic, elastic_object} suggest that heap exploitation remains an ongoing challenge. 

% Prior work
To address heap exploitation effectively, we believe that detecting and capturing heap memory corruption is a key solution. Previously, various research efforts have been made towards this goal, such as Memcheck~\cite{memcheck} and Address Sanitizer~(ASAN)~\cite{address_sanitizer} for comprehensive protection or works that mitigate use-after-free errors~\cite{dangnull,dangsan,markus,dangzero,ffmalloc,cets,freeguard} or detect out-of-bound access~\cite{softbound, in_fat_pointer, low_fat} for partial protection. However, these techniques either introduce significant runtime overhead or offer weak security guarantees. To be specific, MemCheck~\cite{memcheck}, though capable of delivering full heap detection, suffers from a substantial 26x overhead. ASAN and its variant ASAN--, as we will discuss in Section~\ref{sec:eval}, face challenges with false negatives in both temporal and spatial heap error detection~\cite{li2022pacmem}. Other tools like FFmalloc~\cite{ffmalloc}, Delta Pointer~\cite{kroes2018delta}, and Oscar~\cite{dang2017oscar} only offer partial heap protection, limiting their effectiveness.

In this work, we introduce \sys (\textbf{C}ompiler and \textbf{A}llocator-based Heap \textbf{M}emory \textbf{P}rotection), a novel heap sanitizer, for detecting and capturing spatial and temporal heap errors. Unlike previous works~\cite{farkhani2021ptauth, saileshwar2022heapcheck, low_fat, kim2020hardware, in_fat_pointer, woodruff2014cheri, zhang2019bogo, li2022pacmem} that require hardware support, \sys is a software-only tool, consisting of a compiler and a seglist allocator. The compiler instruments the target program to validate the pointer boundary and build point-to relation at runtime. The customized memory allocator tracks memory ranges for each allocation, supports the instrumented instructions, and neutralizes dangling pointers when a memory object is freed. 

In comparison with previous works~\cite{asan_junxu, zhang2021sanrazor}, the key novelty of \sys is mainly manifested in its design that leverages the run-time guarantee to optimize static instrumentation and thus reduces overhead. For example, to capture dangling pointer dereference, previous methods need to check each pointer dereference. In our design, \sys introduces a run-time scheme that could guarantee that no dangling pointer exists and thus allow \sys's compiler to eliminate corresponding pointer dereference checks. Furthermore, as we will detail in Section~\ref{sec:design}, \sys implements its run-time scheme based on the extension of an existing allocator. This allocator has an $O(1)$ computation complexity in pointer validation. As a result, \sys can perform pointer validation more efficiently.

In summary, this paper makes the following contributions.

\begin{itemize}[leftmargin=*, itemsep=1pt, topsep=1pt, partopsep=1pt, parsep=1pt]

\item We present a novel approach \sys that employs a customized allocator and a compiler to safeguard against heap memory corruption. Additionally, we propose optimization strategies to reduce the performance overhead introduced by the instrumentation.

\item We implement \sys by customizing a segregated list allocator -- tcmalloc and building our instrumentation optimization mechanism on top of LLVM 12.0 compiler framework. We open-sourced our prototype of \sys at~\cite{camp_source}.  

\item We conduct a thorough evaluation of \sys using the real-world application Nginx, as well as the SPEC CPU 2006 and 2017 benchmarks, from both security and runtime overhead perspectives. The evaluation compares \sys' performance with other defense solutions offering similar heap protection levels.  

\end{itemize}

The rest of the paper is organized as follows. Section~\ref{sec:bg} introduces the background of memory corruption on the heap as well as heap allocators. Section~\ref{sec:threat_model} discusses the assumptions of our research and the threat model. Section~\ref{sec:design} describes the details of the proposed techniques. Section~\ref{sec:impl} presents our implementation details. Section~\ref{sec:eval} evaluates the security and runtime overhead of our proposed techniques. Section~\ref{sec:dis} provides the discussion of some related issues, followed by the related work in Section~\ref{sec:related}. Finally, we conclude the work in Section~\ref{sec:conclusion}.

\section{Background}
\label{sec:bg}

% In this section, we delve into the background of different types of heap allocators. Besides, we introduce two types of memory corruption errors on the heap -- out-of-bound and use-after-free.  

\subsection{Heap Memory Corruption \& Protection}
% describe how such memory error could happen
In general, there are two main types of heap memory corruption, overflow and use-after-free. In the following, we describe those two types of memory corruption in detail and discuss their protection.

\paragraph{Heap Overflow.} Generally, each heap object has its own memory space. When the access to a heap object exceeds its memory space, a heap overflow happens. One common way to detect heap overflow is to reserve some memory as heap cookies or red zones. Once the magic value in the reserved area is tempered, the heap overflow could be detected~\cite{magic_value}. However, this design naturally has the drawback of being bypassed. For example, the attacker could fail the detection by leaking the heap cookie~\cite{heap_cookie} or overflowing the memory with the red zone intact~\cite{redzone}. An alternative approach is to validate pointers making sure no out-of-bound access happens~\cite{pointer_validation, low_fat, in_fat_pointer}. This approach is effective in general but often introduces non-negligible overhead~\cite{softbound}.

\paragraph{Heap Use-After-Free.} When the memory space of a heap object is freed, the references to the object left become dangling pointers. The program should never deference the dangling pointer. Otherwise, a use-after-free would occur. Researchers have proposed many techniques to detect use-after-free. ASAN~\cite{address_sanitizer} uses shadow memory to record the memory status and instruments every memory access. Accessing a freed object could be detected immediately by checking the shadow memory. Although ASAN's approach only introduces reasonable overhead, its security guarantee is weak where the attacker could overwrite the shadow status by reallocating the freed object. A more effective approach proposed is to never reuse freed memory~\cite{ffmalloc} or delay the free of memory~\cite{delay_mem_free}, so that the attacker would not be able to corrupt freed objects. Other than that, once an object is freed, one could nullify all its existing references~\cite{nullify_references, dangnull, dangsan}, which also prevents use-after-free fundamentally.

\subsection{Heap Memory Allocator}
Heap memory allocators manage dynamic "global" memory and aim for quick allocation/deallocation with minimal memory waste. This summary presents three prevalent types.

Firstly, sequential-fit allocators use a freelist connecting all freed memory objects. When an allocation request is made, the allocator searches the freelist until an adequately sized object is found. It splits larger memory objects, reallocating the excess back into the freelist, and merges neighboring freed objects to minimize fragmentation. Secondly, Segregated List allocators (seglist) manage an array of freelists, each holding freed objects of identical size. Allocation and deallocation require identifying the corresponding freelist for the requested size, avoiding the need for splitting and merging but introducing additional steps. Lastly, the buddy system allocator, similar to seglist, also maintains freelists for varying sizes. When the freelist for a requested size is empty, the allocator splits a larger object to fulfill the request. On deallocation, it reconsolidates the remaining portion back into a larger object.
\section{Assumptions \& Threat Model}
\label{sec:threat_model}
\sys focuses on detecting heap errors, including spatial and temporal heap errors. We assume the target program (written in low-level language) is compiled by \sys and contains at least one heap-based memory vulnerability. As our work focuses on protecting userspace applications, the security of lower-level kernel is out of the scope. In our threat model, the attacker, who is aware of the deployment of \sys, has access to the heap vulnerability and is seeking to exploit it for privilege escalation.
\section{CAMP}
\label{sec:design}
% need an example
% we need to clarify the threat model before this section and justify it in detail in this section
% what is out of scope
% 1. heap memory errors only, i.e. spacial memory error and temporal memory error
%    need to clarify these two concepts in intro and discuss how the state of the art
%    address them. the scope excludes memory errors on stack and global region.
% 2. we do not protect against corruption in external APIs, e.g., libc functions and syscalls, and explain
%    why we and other tools (e.g., KASAN, DangZero, DangNull, etc) do not support this either.
%    also, we should justify that adding support of them is easy with CAMP as long as we can model the memory access
%    of the function, as what we did for memcpy/strcpy in the impl, etc.
% 3. int casting to pointer. this is not allowed because we will lose track of legit pointers. We ensure
%    this property through compiler: once compiler detects such a case, it will refuse to compile.

% something that is extended from background

\begin{figure}
% \vspace{-4ex}
\begin{lstlisting}[caption={A toy vulnerable example.},label={lst:toy_program}]
void main() {
    char *buf = malloc(16);
    buf[32] = 'x';
    free(buf);
    buf[1] = 'y';
}
\end{lstlisting}
\vspace{-2ex}

\begin{lstlisting}[caption={The toy program with CAMP's protection.},label={lst:protected_toy_program}]
void main() {
    char *buf = malloc(16);
    __escape(&buf, buf);
    __check_range(buf, &buf[32], sizeof(char));
    buf[32] = 'x';
    // free buf, which neutralizes the dangling pointer stored in &buf
    free(buf);
    __check_range(buf, &buf[1], sizeof(char));
    buf[1] = 'y';
}
\end{lstlisting}
% \vspace{-4ex}
\vspace{-2ex}
\end{figure}

% In this section, we first give an example of how \sys protects a vulnerable program, then introduce its design details.

\subsection{A vulnerable toy program}
List~\ref{lst:toy_program} illustrates a toy program that contains two heap memory corruption vulnerabilities. Specifically, the program allocates 16 bytes of memory (line~2), then accesses the memory object at index 32 which exceeds the boundary of the memory range (line~3), causing a heap memory overflow. It is noted that this heap overflow could not be detected by ASAN because the overflow skips the red zone. In line~4, the memory object is freed, which makes the pointer \code{buf} a dangling pointer. After this, the dangling pointer is dereferenced (line~5), resulting in a use-after-free memory corruption.

\subsection{CAMP's Protection}
\label{sec:design:example}
% illustrate how the protection works
Briefly speaking, \sys instruments the program to detect memory corruption and prevent exploitation. List~\ref{lst:protected_toy_program} illustrates the toy program protected by \sys. 
In the following, we describe how \sys protects the vulnerable toy program.
% checks on poiner arith and assignment
% discuss oob pointers from libc calls, or syscalls. supporting those requires a specific model of the function
% we define our program model that does not allow int to pointer cast.
% how do we prove that an oob must have an oob pointer?

\paragraph{Pointer Validation.} \sys detects heap overflow by validating result pointers from pointer arithmetic, making sure no out-of-bound pointer is generated. This is achieved by adding a check instruction at the site of the pointer arithmetic to query the runtime and verify that the result pointer is within the range of the base pointer. As demonstrated in List~\ref{lst:protected_toy_program}, pointers \code{&buf[32]} and \code{&buf[1]} are generated based on the buffer \code{buf} in lines 5 and 9, respectively. \sys automatically instruments check instructions in lines 4 and 8 to validate these pointers. If the runtime detection determines that \code{&buf[32]} is an out-of-bound pointer, \sys will abort the execution to prevent exploitation. Note that the query requires \sys to maintain awareness of each memory allocation and its associated memory range, which is recorded during runtime for each heap allocation (line 2).
% As \sys does not utilize the red zone, it could effectively catch non-linear out-of-bound access which ASAN is not capable of.

\paragraph{Neutralizing Dangling Pointers.} \sys prevents use-after-free by neutralizing dangling pointers. At runtime, \sys constructs the point-to relation by instrumenting the program. When a memory object is freed, \sys could look up the built point-to relation and identify the dangling pointers to the freed memory. By neutralizing those dangling pointers, use-after-free access is no longer possible. Copying pointers (i.e., \textit{pointer escapes} ~\cite{carat_native}) is tracked to build the point-to relation. For example, the program copies the heap pointer to the variable \code{buf} (line~2 of List~\ref{lst:protected_toy_program}), which is a pointer escape and \sys instruments an escape tracking instruction after that (line~3). The escape tracking instruction takes as input the address of the variable and the pointer, annotating which address contains a reference to the memory allocation. After this, the program frees the memory (line~7). Inside the free, \sys will identify the existing dangling pointers to the freed memory and neutralize them as non-congenial. As a result, the variable \code{buf} no longer references the freed memory and the program crashes when it is dereferenced (line~9).

\begin{figure}[t]
\centering
% \vspace{-4ex}
\includegraphics[width=0.45\textwidth]{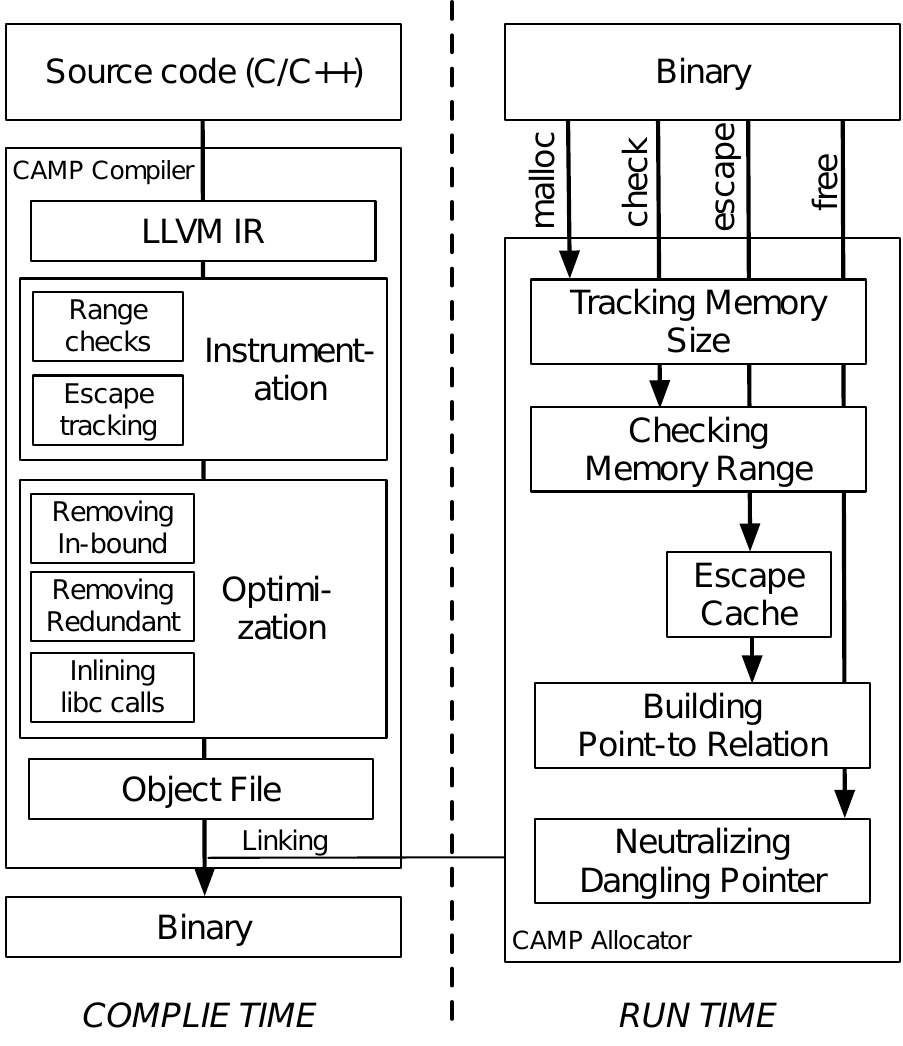}
\caption{The design overview of CAMP.}
\vspace{-3ex}
\label{fig:overview}
\end{figure}

% \begin{figure}[t]
% \centering
% \includegraphics[width=0.45\textwidth]{fig/allocator.eps}
% \caption{Overview}
% \label{fig:tcmalloc}
% \end{figure}

\subsection{Design Overview}
Figure~\ref{fig:overview} shows the core components of \sys, which consists of a compiler and a memory allocator. The \sys compiler, which is built on top of LLVM, takes the source code as inputs and outputs binaries linked with \sys allocator. During the compile time, the compiler first translates the source code into LLVM IR and then instruments the range checking and escape tracking instruction to defend against heap memory corruption. After this, the compiler applies several novel compiler optimizations to reduce the overhead of protection, without sacrificing security guarantees.
At run-time, \sys's allocator handles the heap memory allocation/deallocation request. In addition, it provides support for instrumented instructions. Specifically, the allocator tracks the memory range for each allocation, so that the allocator could validate the bound information of pointers for each range checking query.
The allocator also handles the escape tracking instruction to build the point-to relation. With this,
whenever a memory object is freed, it is capable of neutralizing the corresponding dangling pointers, preventing the UAF access from dangling pointers.

\subsection{Compiler Instrumentation}
\label{sec:compiler_inst}
\paragraph{Instrumenting Range Checking.} As our program model does not allow casting integers to pointers, initial pointers are all from either explicit memory allocation (i.e., malloc), or taking the address from global or stack variables.~\cite{softbound} %softbound
Those pointers then are used through \textit{pointer arithmetic} to create new pointers, then to access memory. The range checking in \sys is to ensure that all pointers after arithmetic are still in-bound, therefore, preventing heap overflow. As is shown in List~\ref{lst:toy_program}, the range checking takes three arguments, which are the base pointer $src$, the result pointer $dst$ of the arithmetic, and $dst$'s type size $size$, respectively. The run-time will assert that the memory ranging from $dst$ to $dst+size$ is within the memory range of $src$.

It is noted \sys only protects heap memory, so validating non-heap memory pointer is redundant. To save unnecessary validation, \sys performs dataflow and alias analysis on the LLVM IR to determine the point-to relation of pointers. If a pointer could be determined at compile time that it does not point to the heap, \sys will not instrument range checking for it.
% justify the advantage of this instrumentation
% will not skip the red zone
% light compared to memory load and store checks
\paragraph{Instrumenting Escape Tracking.} The inserted escape tracking allows \sys to build the point-to relation of memory objects at run-time. We follow a similar approach introduced in CARAT~\cite{carat_native, carat_cake} and DangNull~\cite{dangnull} to insert the tracking after \textit{pointer escapes} (i.e., copying pointers). As is described in Section~\ref{sec:design:example}, the tracking takes the copied pointer and its stored address as arguments. Different from CARAT and DangNull, which instrument all the potential pointer escapes, \sys skips pointer escapes if the pointer of which is determined not to reference the heap at compile time. As the goal of \sys is protecting heap memory error, skipping those non-heap point-to relations does not jeopardize the security but helps \sys obtain better performance.
% instrument point-to instructions
% this belongs to implementation. Specifically, the \sys compiler goes through all the LLVM IR, if a \code{store} instruction indicates an assignment of a pointer to memory, the compiler will insert an escape tracking instruction.

\subsection{Runtime Support}
% support query of slot size given a heap address
% support maintaining the point-to relationship between memory slots
% justify the advantage of the design of our allocator v.s. dangnull
The runtime of \sys provides underlying support of \sys's instrumentation. 
The performance of executing inserted instruction is critical to the overall performance of \sys.
Assuming that a program has $m$ pieces of memory allocated, and \sys maintain the record of the allocated memory in a linked list. In the worst case, performing a single range checking will cause $O(m)$ time complexity, which apparently would introduce unattainable runtime overhead. Likewise, a dummy design that records $n$ pointer escapes in a linked list would introduce $O(n)$ time complexity when freeing an object.
In the following, we describe how our design cooperates with the allocator to provide fast runtime support.

\paragraph{Seglist Allocator.} As we mentioned in Section~\ref{sec:bg}, a Seglist Allocator uses different free lists for different sizes of memory objects. For example, tcmalloc~\cite{tcmalloc}, a Segregated List Allocator developed by Google, uses \textit{span} as the basic memory management unit. Each span manages a size class of memory objects on several continuous memory pages. The spans are stored in a page table where the page is used as the key. Whenever allocating/deallocating a memory object, tcmalloc could find its span, and then retrieves the freelist with constant time complexity.

\sys's allocator leverages the design of the Segregated List Allocator to provide fast runtime support. For each span, \sys records the size of memory objects as one of the metadata. Since the seglist allocator splits the memory page equally into objects, given a heap pointer denoted as $ptr$, we could first calculate the index of the object with:

$idx = (ptr - page\_base) / size$

\noindent{}where $idx$ is the object index to the span, $page\_base$ represents the start address of the page in the span, and $size$ refers to the size of the object.
With this information, the lower bound of the memory range can be identified as $page\_base + idx * size$ and the upper bound as $page\_base + (idx + 1) * size$.
This straightforward approach results in constant time complexity for pointer validation.

\paragraph{Maintaining Point-to Relation.} 
\sys differs from DangNull~\cite{dangnull} and CARAT~\cite{carat_native, carat_cake} in its approach to encoding the point-to relation. While DangNull and CARAT use a red-black tree structure, which optimizes the time complexity of finding a relation to $O(log N)$, \sys integrates the point-to information into the seglist allocator, optimizing cost to constant time complexity.
Specifically, \sys's seglist allocator maintains an escape table for each span, which is a map of object indices to their escape lists. The escape lists are linked-list structures that chain corresponding escapes to their dedicated objects. When an escape tracking call is made, \sys calculates the memory object's index in the span, retrieves its escape list from the table, and inserts a record into the list.
Upon a memory object is freed, \sys's allocator checks its escape list and neutralizes all the existing dangling pointers to the memory object. 
% Compared with DangNull, our design has less memory overhead for maintaining the point-to relation. For each point-to relation, our approach requires only 16 bytes to maintain it since the escape list table is generally small, whereas DangNull requires 24 bytes of memory. What is more, our design of constructing point-to relations and neutralizing dangling pointers provide constant time complexity. DangNull introduces $O(logN)$ time complexity, where $N$ is the number of the pointer escapes. As we will show in Section~\ref{sec:eval}, \sys, on average, outperforms \fixme{xx} runtime performance and reduces \fixme{xx} memory overhead compared with DangNull.

To further boost the overall performance of \sys, we design a cache mechanism for maintaining the point-to relation. New point-to relations are temporarily stored in the cache until it becomes full, at which point the records are transferred to the allocator in a batch, while skipping any duplicates. This cache design boosts runtime speed and reduces memory overhead, particularly in scenarios where the program operates repeatedly in the same block and creates similar point-to relations.
% need a graph explaining the whole arch

% need to explain the complexity of the runtime, and compare with others

\begin{figure}
\begin{lstlisting}[caption={An example of optimizing structure field access checks.},label={lst:opt_structure}]
struct obj {
    int a;
    int b;
};
struct obj* bar() {
    // type-casting from void* to obj*
    struct obj *o = malloc(sizeof(struct obj));
    __check_range(o, o, sizeof(struct obj));
    ...
}
int foo(struct obj *ptr) {
    __check_range(ptr, &ptr->a, sizeof(ptr->a));
    ptr->a = 1;
    __check_range(ptr, &ptr->b, sizeof(ptr->b));
    ptr->b = 2;
}
\end{lstlisting}
\label{lst:opt_structure}
\vspace{-2ex}
\end{figure}

\subsection{Compilation Optimization}
% \subsubsection{ Optimization}
One of the \sys's characteristics is that it leaves a large room for potential optimization during the compilation time. This characteristic could boost performance significantly and, at the same time, preserve security. In the following, we describe three optimization algorithms we design for \sys. 
% But, as we will discuss in Section~\ref{}, other optimizations could also be added to further improve \sys. We leave the exploration of other compilation optimizations as part of our future research.

\begin{algorithm}[t]
\small
\textbf{Input:} A function $F$ \;
\textbf{Output:} A set of pointer to be validated $S$ \;
\textbf{Initialize:} $NewPointerSet$ = \textbf{\textit{getNewPointerSet}}\{$F$\} \; $In = Out = changeSet = dict()$ \;
\ForEach{ptr $\in$ NewPointerSet}{
    \If{pointerMapbase[\textbf{base}(ptr)] == NULL} {
        $pointerMapbase[\textit{\textbf{base}}(ptr)] = set()$ \;
    }
    $pointerMap$[$\textbf{\textit{base}}(ptr)$]$.$add$(ptr)$ \;
}
$In = pointerMap$ \;
$changeSet = In - Out$ \;
\While{ changeSet $\neq$ $\emptyset$}{
    \ForEach{ key, val $\in$ In}{
        \If{$\exists$ p, p' $\in$ val \AND \textbf{RedundantPair}(p, p')} {
            $val.remove(p')$ \;
            $val.p.offset = \textbf{MAX}(p.offset,  p'.offset)$ \;
            $Out[key] = val$ \;
            \textbf{break};
        } \Else {
            $Out[key] = val$ \;
        }
    }
    $changeSet = In - Out$ \;
    $In = Out$ \;
}
\ForEach{ key, val $\in$ Out}{
$S.add(val)$ \;
}
\caption{Removing Redundant Validation}
\label{algo:removing_redundant}
\end{algorithm}

\paragraph{Optimizing range checks with type information.} 
% ensure the memory slot is large enough for its type
% the exception is the elastic object

% With this, all access to the slot field is optimized
% array are treated as structures at runtime to apply this optimization
A naive design against out-of-bound access is to have range checks on every pointer arithmetic to ensure that pointers do not exceed memory bounds. As an example, consider the function \code{foo} in List~\ref{lst:opt_structure}. The variable \code{ptr} in lines 13 and 15 involves two pointer arithmetics, and \sys must insert range checks for the result pointers (lines 12 and 14) for security purposes.
An optimization strategy is that if the compiler knows the pointer is in-bound, the validation of which could be removed to improve the performance. However, the compiler typically lacks information about the memory range of a pointer, making it challenging to determine which pointers are within bounds.

To apply the optimization, we use type information to determine the memory range of a pointer during the compilation process. This involves not only validating pointer arithmetic, but also validating type-casting operations to ensure the memory space referenced by a typed pointer is adequate for its corresponding type.
For example, in the code shown in List~\ref{lst:opt_structure}, the function \code{bar} allocates a new object \code{obj} (line 7). The return type of \code{malloc} is \code{void*}, not \code{struct obj*}, so the compiler inserts a type-casting instruction, after which \sys inserts a range check to ensure the memory space is sufficient to hold the structure \code{obj} (line 8). 
With this type-casting validation, the compiler can safely infer that the memory space of a typed pointer is at least its type size. As a result, the compiler can conclude that pointers referring to the structure field are in-bound. Therefore, \sys can remove the range checks in lines 12 and 14. Further, The compiler can also guarantee that the memory of pointer \code{o} (line 6) is at least the size of its type, so the range check in line 7 can also be optimized and eliminated.

% As a result, all the range checks in List~\ref{lst:opt_structure} could be removed without jeopardizing \sys's security guarantee.

% enphasize that this optimization could be applied to ASAN, because it will break the security model of ASAN's design
The optimization is made possible by the unique design of security checks in \sys, which eliminates the risk of accessing to a dangling pointer, such as \code{ptr} in function \code{foo}. This allows for aggressive optimization of range checks for field pointers without the concern of casing a false-negative for use-after-free vulnerability.
As we will show in Section~\ref{sec:eval}, the proposed optimization approach above dramatically improves \sys's runtime performance, especially for programs that contain pointers associated with types. It should be noted that for pointers that have no type information (e.g., \code{void*, char *}) or their type sizes could not be determined at compilation time (e.g., elastic objects~\cite{elastic_object}), we do not apply the optimization above.
As ASAN's security design cannot guarantee those conditions, applying this optimization to it may result in use-after-free and out-of-bound access.

% This optimization approach dramatically improves \sys's runtime performance, especially for programs that use types more, as we will show in Section~\ref{sec:eval}. Noted that for pointers that have no type information (e.g., \code{void*, char *}) or whose type size could not be determined at compile-time (e.g., elastic objects~\cite{}), the optimization does not apply. 

% \begin{algorithm}[t]
% \small
% \textbf{Input:} pointerSet \;
% \textbf{Output:} pair \;
% \textbf{Initialize:} pair = $\emptyset$ \;
% \ForEach{ptr $\in$ pointerSet}{
%     \ForEach{ptr' $in$ pointerSet}{
%         \If{ptr == ptr'} {
%             \textbf{continue} \;
%         }
%         \If{\textbf{isConst}(ptr.offset) $\AND$ \textbf{isConst}(ptr'.offset)} {
%             \If{\textbf{dominate(ptr, ptr')} $\OR$ \textbf{postDominate(ptr, ptr'}} {
%                 pair = (ptr, ptr') \;
%                 break;
%             }
            
%         }
%     }
% }
% \caption{algo of findPair}
% \label{worklist}
% \end{algorithm}

\begin{figure}
% \vspace{-4ex}
\begin{lstlisting}[caption={Example codes of applying eliminating redundant optimization.},label={lst:redundant_opt}]
struct obj {
    char *mem;
};
void foo(struct obj *ptr, bool flag) {
    __check_range(ptr->mem, &ptr->mem[0x100], 1);
    ptr->mem[0x100] = 'x';
    if (flag) {
        ptr->mem[0x30] = 'y';
        ptr->mem[0x1] = 'y';
    }
    ptr->mem[0x1] = 'z';
    ptr->mem++;
}
\end{lstlisting}
\vspace{-2ex}
\end{figure}

% argue this is a very secure coding fashion that is used in the community:
% all the test cases in the eval align with this fashion except PHP which contains
% only one site of casting a small memory slot to a large type. We send the PR and
% the developer acked the poor coding and accepted our pr very quickly.
\paragraph{Removing Redundant Instructions.} 
% in the same dominate path, only check the least range
% for the same pointer, at the same dominate tree, check the least pointer.
% fix-point algo
% compare it with ASAN--, e.g., cheaper, and secure: with only one check, obtain better security.
% 1. construct a map of src pointer and its guard
% 2. for each ele in the map, if one new pointer dominates or post-dominates another or others, remove the guard at dominatee and use the biggest offset for guard at dominator. if one new pointer post dominate another.
% 3. update the map
% 4. if 2 found a satisfied case, then repeat 2, otherwise exit. (fix-point)
% loop ideally satisfies this algo if we can unroll all the loops. But unrolling all loop is difficult, therefore we treat loops as loop invariant.
% escape redundant: a->b++;
A redundant instruction in this work refers to a range checking that validates pointers that have been validated or an escape tracking that builds point-to relation that has been recorded. The optimization opportunity for this is to remove the redundant ones to obtain better performance.

Intuition suggests that if two range checks validate the same result pointers from pointer arithmetic, one of them could be removed. For example, in List~\ref{lst:redundant_opt}, the pointer \code{&ptr->mem[1]} in line~9 and 11 aliases. We could simply remove the validation for \code{&ptr->mem[1]} in line~9. Further, \sys's security design allows merging several pointer validations into one validation. Initially, \sys needs to validate the result pointers \code{&ptr->mem[0x100]} (line~6), \code{&ptr->mem[0x30]} (line~8), and \code{&ptr->mem[0x1]} (line~11), respectively. However, if pointer \code{&ptr->mem[0x100]} is in-bound, the other two pointers must be in-bound as well. Therefore, we could remove their validations and move the validation of \code{&ptr->mem[0x100]} to line~5, as is illustrated in List~\ref{lst:redundant_opt}. Formally, given two result pointers $ptr1$ and $ptr2$ from the same base pointer $ptr$, their validation can be merged if the following function returns True.
\begin{algorithmic}
\Function{RedundantPair}{$ptr1, ptr2$}
\State \textbf{if} \textit{ptr1.offset >= ptr2.offset} \textbf{then} 
\State \textbf{    } \textbf{if} \textit{dominate(ptr1, ptr2) \OR}
\State \textbf{} \textbf{} \textbf{} \textbf{} \textbf{} \textit{post-dominate(ptr1, ptr2)} \textbf{then}
\State \textbf{} \textbf{} \textbf{} \textbf{} \textbf{} \textbf{} \textbf{} \Return \textbf{True}
\State \Return \textbf{False}
% \Return something
\EndFunction
\end{algorithmic}

% \textit{ptr1.offset >= ptr2.offset},
% \item[1]
% and
% \item[1]
% \textit{dominate(ptr1, ptr2) || post-dominate(ptr1, ptr2)}.

\noindent{}The $offset$ in the first condition represents the maximum access offset from the base pointer. As such, the condition requires $ptr2$ to be in the range of $[ptr, ptr1]$. The second condition ensures the redundancy of validation, where the two validations will be executed together. We follow Algorithm~\ref{algo:removing_redundant} to remove redundant pointer validation. The algorithm takes as input a function $F$, and outputs a set of pointers to be validated $S$. We first collect all the result pointers (line~3) and categorize them into a map according to their base pointers (line~5 to 8), where the key is the base pointer, value is the set of result pointers. Then we follow the fix-point algorithm~\cite{fixed_point} to apply the optimization. In each iteration, we go through the element in the map (line~12). if we find two pointers that satisfy the condition of the redundancy (line~13), we remove the latter one (line~14) and adjust the remaining one's offset with their maximum value (line~15). After this, we update the output of iteration (line~16) and exit the loop to start the next iteration. Noted that if no redundancy pair is found, the output will be just the input of the iteration (line~19). When there is no redundancy pair left, meaning that the fixed point is reached. Then we collect the remaining pointers from the $out$ to $S$ (line~22 to 23). List~\ref{lst:redundant_opt} showed \sys's instrumentation after applying this optimization, where only the validation in line~5 is preserved, but the security guarantee remains. Note that applying this optimization to redzone-based protection (e.g., ASAN) may remove out-of-bound access checks. Assuming that \code{ptr->mem} is a 0x20 bytes memory chunk with 0x10 bytes red zone, normally the overflow will be detected through the check for \code{ptr->mem[0x30]} in line~8. However, applying the proposed optimization will remove this valid check but keep the one for \code{ptr->mem[0x100]}, which skips the red zone and miss the overflow detection, thereby leading to false negatives.

In line 12 of List~\ref{lst:redundant_opt}, the pointer \code{ptr->mem} is self-updated. If we break this statement down, there are three operations involved. First, the program retrieves the pointer stored in address \code{&ptr->mem}. Then, it creates a new pointer based on the retrieved pointer, where a range checking will be inserted. This check guarantees the newly created pointer references to the same memory object as the old one does. Then, the new pointer is copied into address \code{&ptr->mem}, after which \sys will insert an escape tracking to the pointer copying. One key observation is that the address \code{&ptr->mem} should have been initialized somewhere before, which means such a point-to relation must have been recorded. Because the memory object of the new pointer is not changed, recording the same point-to relation is redundant. Therefore, we could optimize this escape tracking for performance without sacrificing precision. Our optimization strategy is to identify those escape pairs that are doing self-updating.

\paragraph{Merging Runtime Calls.}
% fixme:
% if a pointer the same pointer is being checked more than once, we move the check in context.
% add details of how the alias is determined
% add details of how the security is preserved
Ideally, for the same memory pointer (including alias), the validation can be optimized with the aforementioned approach into a single range check. However, if the pointer arithmetic is dynamic, where the result of pointer arithmetic cannot be determined at the compilation time, \sys has to instrument different range checks for them.
List~\ref{lst:merge_calls} illustrates such example codes. Line~5 and 7 access the same array \code{ptr} that does not have type information. Besides, the access is dynamically based on the runtime value \code{i} and \code{j}, making the optimization of removing redundancy not applicable.
To this end, \sys needs to instrument both Line~5 and 7 respectively.
Each time executing a range check, \sys has to switch its context into the library, query the memory range, and then validate the pointer. This process is time-consuming.
One strategy to optimize this is that we could merge the range checks since they share the same base pointer. 
This can reduce the cost of constantly switching contexts and save the time of querying the memory range. 

To do this, we first follow the same approach in Algorithm~\ref{algo:removing_redundant} of constructing a \textit{pointer map} where the pointer arithmetic with the same base pointer is categorized into the same group. Then, for each group, we go through the CFG of the function and find their nearest \textbf{\textit{dominator}} instruction, where a range query is inserted to initialize the memory chunk's range variables. After this, the original range check then is replaced with assertion to ensure the boundary.
List~\ref{lst:merge_calls} also shows the result after applying this strategy. Line~3 queries the memory range of \code{ptr} and initializes the memory ranges into variables \code{start} and \code{end}. After this, validation for \code{&ptr[i]} and \code{&ptr[j]} is done through two assertions in line~4 and line~6.
\begin{figure}
\begin{lstlisting}[caption={Example codes of applying merging runtime calls optimization.},label={lst:merge_calls}]
void foo(char *ptr, int i, int j) {
    unsigned int start, end;
    __get_range(ptr, &start, &end);
    assert(&ptr[i]>=start && &ptr[i]+1<end);
    ptr[i] = 'x';
    assert(&ptr[j]>=start && &ptr[j]+1<end);
    ptr[j] = 'y';    
}
\end{lstlisting}
\vspace{-2ex}
\end{figure}
\section{Implementation}
\label{sec:impl}
In this section, we describe our implementation of \sys's compiler and the allocator. 
% The source code of \sys could be found at ~\cite{camp_source}.

\paragraph{CAMP Compiler.}
\sys's compiler is built on top of LLVM 12 compiler framework. We implement the instrumentation and the optimization in an LLVM pass, loadable by clang.

% shrunk version
To defend against heap overflow, it instruments all pointer arithmetic and type-casting instructions. In the context of pointer arithmetic, the compiler collects the \code{getelementptr} instruction from the LLVM IR, which represents the sole pointer arithmetic instruction as \sys prohibits casting integers to pointers. \sys adds a range checking instruction after every \code{getelementptr} instruction, with three inputs: the base pointer, the result pointer, and the type size of the result pointer. The \code{getelementptr} instruction's pointer operand, result value, and type size serve as these inputs respectively. To determine if the \code{source} operand refers to the heap, \sys backtracks it following LLVM's SSA to identify its origin. If the source is a stack or global variable, the system presumes it doesn't refer to the heap, skipping the \code{getelementptr} instruction's instrumentation. The compiler also instruments the \code{bitcast} instruction, which symbolizes type-casting in LLVM. For each type-casting instruction, \sys adds a checking instruction with two inputs - the result pointer and its type size - to ensure adequate memory for the object.

Following CARAT's approach~\cite{carat_native, carat_cake} to tracking escapes, \sys instruments store instructions if their value operand type is a pointer. If the escape is on heap memory, the compiler inserts an escape tracking CALL instruction before the store instruction.

\paragraph{Memory Allocator.} The allocator is built on tcmalloc~\cite{tcmalloc}. Tcmalloc maintains a page table mapping page addresses to a \code{span}. Notably, a span can handle multiple continuous memory pages for larger objects. To ensure any heap pointer can find its \code{span}, we register every memory page used by tcmalloc in the page map. Each span is supplemented with two metadata to facilitate \sys's runtime checks: the object size and a reference to the escape pointer array containing linked escapes to the objects it manages. For each page, we compress its span's start address and size class into an 8-byte unit and map it into the \textit{size class map}. With this design, \sys can retrieve the necessary 
data and validate pointers in constant time. For unrecorded spans, \sys resorts to the original routine of retrieving the \code{span}.

To accommodate coding styles that use out-of-bound pointers as memory boundaries, we reserve one-byte memory at the end of each allocation, ensuring such pointers remain in-bound. Each pointer escape prompts \sys to create a record containing the pointer's location, stored in the span of the referenced object. Linked escape records for the same object are created, and all related records are freed when a memory object is freed. During this process, \sys neutralizes any pointers still referencing the object. \sys features a temporal escape array cache mechanism. Every time a new escape track is invoked, \sys checks the array for an identical record before appending. Once full, all records are committed to the span and the array cleared, ensuring all dangling pointers are neutralized.
\begin{table*}[t]\small
    \setlength\tabcolsep{5pt}
    \centering
    \begin{tabular}{ccc}
       \toprule
        \normalsize{\textbf{CWE (number)}} & 
        \makecell[c]{\normalsize{\textbf{Good Test}} \\
        \scriptsize{(Selected/Total/Passed)}}
        & 
        \makecell[c]{\normalsize{\textbf{Bad Test}} \\
        \scriptsize{(Selected/Total/Passed)}} \\
       \midrule
       \makecell[c]{Buffer 
        Overflow(122)} & 
        3870/3870/3870
        & 
        2308/3870/2308 \\
        Double Free(415) & 820/820/820 & 820/820/820 \\
        Use After Free(416) & 394/394/394 & 288/288/288 \\
        Invalid Free(761) & 288/288/288 & 288/288/288 \\
       \bottomrule
    \end{tabular}
    \caption{Security evaluation of CAMP on Juliet Test Suite.}
    % \vspace{-2ex}
    \label{tab:juliet}
\end{table*}

\begin{table*}[t]\small
    % \vspace{-20pt}
    \tabcolsep=4pt
    \centering
        \begin{tabular}{llclllllc}
         \toprule
            \normalsize{\textbf{CVE/Issue ID}} & 
            \normalsize{\textbf{Application}} & 
            \normalsize{\textbf{Bug Type}} &
            \normalsize{\textbf{CAMP}} &
            \normalsize{\textbf{ASAN~\code{--}}} &
            \normalsize{\textbf{Memcheck}} &
            \normalsize{\textbf{DangNull}} &
            \normalsize{\textbf{MarkUs}} &
            \normalsize{\textbf{Delta pointer}}\\
         \midrule
           CVE-2015-3205 & libmimedir & Use-After-Free & \multicolumn{1}{c}{\ding{52}} & \multicolumn{1}{c}{\ding{52}} & \multicolumn{1}{c}{\ding{52}} & \multicolumn{1}{c}{\ding{52}} & \multicolumn{1}{c}{\ding{52}} & \multicolumn{1}{c}{/}\\
           
           CVE-2015-2787 & PHP 5.6.5 & Use-After-Free & \multicolumn{1}{c}{\ding{52}} &
           \multicolumn{1}{c}{\ding{52}} & \multicolumn{1}{c}{\ding{52}} & \multicolumn{1}{c}{\ding{54}} & \multicolumn{1}{c}{\ding{54}} & \multicolumn{1}{c}{/}\\
           
           CVE-2015-6835 & PHP 5.4.44 & Use-After-Free & \multicolumn{1}{c}{\ding{52}} & 
           \multicolumn{1}{c}{\ding{52}} & \multicolumn{1}{c}{\ding{52}} & \multicolumn{1}{c}{\ding{52}} & \multicolumn{1}{c}{\ding{54}} & \multicolumn{1}{c}{/}\\
           
           CVE-2016-5773 & PHP 7.0.7 & Use-After-Free & \multicolumn{1}{c}{\ding{52}} &
           \multicolumn{1}{c}{\ding{52}} & \multicolumn{1}{c}{\ding{52}} & \multicolumn{1}{c}{\ding{52}} & \multicolumn{1}{c}{\ding{54}} & \multicolumn{1}{c}{/}\\
           
           Issue-3515~\cite{mruby_issue_3515} & mruby & Use-After-Free & \multicolumn{1}{c}{\ding{52}} & 
           \multicolumn{1}{c}{\ding{52}} & \multicolumn{1}{c}{\ding{52}} & \multicolumn{1}{c}{Build Fail} & \multicolumn{1}{c}{\ding{54}} & \multicolumn{1}{c}{/}\\
           
           CVE-2020-6838 & mruby & Use-After-Free & \multicolumn{1}{c}{\ding{52}} & \multicolumn{1}{c}{\ding{52}} & \multicolumn{1}{c}{\ding{52}} & \multicolumn{1}{c}{Build Fail} & \multicolumn{1}{c}{\ding{54}} & \multicolumn{1}{c}{/}\\
           
           CVE-2021-44964 & Lua & Use-After-Free & \multicolumn{1}{c}{\ding{52}} & \multicolumn{1}{c}{\ding{52}} & \multicolumn{1}{c}{\ding{52}} & \multicolumn{1}{c}{Build Fail} & \multicolumn{1}{c}{\ding{52}} & \multicolumn{1}{c}{/}\\
           
           CVE-2020-21688 & FFmpeg & Use-After-Free & \multicolumn{1}{c}{\ding{52}} & \multicolumn{1}{c}{\ding{52}}& \multicolumn{1}{c}{\ding{52}} & \multicolumn{1}{c}{\ding{54}} & \multicolumn{1}{c}{\ding{52}} & \multicolumn{1}{c}{/}\\
           
           CVE-2021-33468 & yasm & Use-After-Free & \multicolumn{1}{c}{\ding{52}} & \multicolumn{1}{c}{\ding{52}} & \multicolumn{1}{c}{\ding{52}} & \multicolumn{1}{c}{\ding{52}} & \multicolumn{1}{c}{\ding{52}} & \multicolumn{1}{c}{/}\\
           
           CVE-2020-24978 & nasm & Use-After-Free & \multicolumn{1}{c}{\ding{52}} & \multicolumn{1}{c}{\ding{52}} & \multicolumn{1}{c}{\ding{52}} & \multicolumn{1}{c}{\ding{54}} & \multicolumn{1}{c}{\ding{52}} & \multicolumn{1}{c}{/}\\
           
           Issue-1325664~\cite{chrome_issue_1325664} & Chrome & Use-After-Free & \multicolumn{1}{c}{\ding{52}} & \multicolumn{1}{c}{\ding{52}} & \multicolumn{1}{c}{\ding{52}} & \multicolumn{1}{c}{Build Fail} & \multicolumn{1}{c}{\ding{54}} & \multicolumn{1}{c}{/}\\
           
           CVE-2022-43286 & Nginx & Use-After-Free & \multicolumn{1}{c}{\ding{52}} & \multicolumn{1}{c}{\ding{52}} & \multicolumn{1}{c}{\ding{52}} & \multicolumn{1}{c}{\ding{54}} & \multicolumn{1}{c}{\ding{52}} & \multicolumn{1}{c}{/}\\
           
           CVE-2019-16165 & cflow & Use-After-Free & \multicolumn{1}{c}{\ding{52}} & \multicolumn{1}{c}{\ding{52}} & \multicolumn{1}{c}{\ding{52}} & \multicolumn{1}{c}{\ding{54}} & \multicolumn{1}{c}{\ding{52}} & \multicolumn{1}{c}{/}\\
           
           CVE-2021-4187 & vim & Use-After-Free & \multicolumn{1}{c}{\ding{52}} & \multicolumn{1}{c}{\ding{52}} & \multicolumn{1}{c}{\ding{52}} & \multicolumn{1}{c}{\ding{54}} & \multicolumn{1}{c}{\ding{52}} & \multicolumn{1}{c}{/}\\
           
           % CTF Challenge & ghostparty & Use-After-Free & Exploit Failed \& Exception Happen \\
           % CTF Challenge & Use-After-Free & Use-After-Free & \multicolumn{1}{c}{\ding{52}}n \\
           % CTF Challenge & Unsubscriptions-Are-Free & Use-After-Free & \multicolumn{1}{c}{\ding{52}} \\
           CVE-2022-0891 & libtiff & Heap Overflow & \multicolumn{1}{c}{\ding{52}} & \multicolumn{1}{c}{\ding{52}} & \multicolumn{1}{c}{\ding{52}} & \multicolumn{1}{c}{/} & \multicolumn{1}{c}{/} & \multicolumn{1}{c}{\ding{52}}\\
           
           CVE-2022-0924 & libtiff & Heap Overflow & \multicolumn{1}{c}{\ding{52}} & \multicolumn{1}{c}{\ding{52}} & \multicolumn{1}{c}{\ding{52}} & \multicolumn{1}{c}{/} & \multicolumn{1}{c}{/} & \multicolumn{1}{c}{\ding{52}}\\
           
           CVE-2020-19131 & libtiff & Heap Overflow & \multicolumn{1}{c}{\ding{52}} & \multicolumn{1}{c}{\ding{52}} & \multicolumn{1}{c}{\ding{52}} & \multicolumn{1}{c}{/} & \multicolumn{1}{c}{/} &  \multicolumn{1}{c}{\ding{52}}\\
           
           CVE-2020-19144 & libtiff & Heap Overflow & \multicolumn{1}{c}{\ding{52}} & \multicolumn{1}{c}{\ding{52}} & \multicolumn{1}{c}{\ding{52}} & \multicolumn{1}{c}{/} & \multicolumn{1}{c}{/}& \multicolumn{1}{c}{\ding{52}}\\
           
           CVE-2021-4214 & libpng & Heap Overflow & \multicolumn{1}{c}{\ding{52}} & \multicolumn{1}{c}{\ding{52}} & \multicolumn{1}{c}{\ding{52}} & \multicolumn{1}{c}{/} & \multicolumn{1}{c}{/} & \multicolumn{1}{c}{Build Fail}\\
           
           % CVE-2022-2566 & FFmpeg & Heap Overflow & \multicolumn{1}{c}{\ding{52}} \\
           
           CVE-2021-3156 & sudo & Heap Overflow & Run Well  & \multicolumn{1}{c}{\ding{52}} &  \multicolumn{1}{c}{\ding{52}} &  \multicolumn{1}{c}{/} &   \multicolumn{1}{c}{/} &  \multicolumn{1}{c}{Build Fail}\\
           
           CVE-2018-20330 & libjpeg-turbo & Heap Overflow & \multicolumn{1}{c}{\ding{52}} & \multicolumn{1}{c}{\ding{52}} & \multicolumn{1}{c}{\ding{52}} & \multicolumn{1}{c}{/} & \multicolumn{1}{c}{/} & \multicolumn{1}{c}{\ding{52}} \\
           
           CVE-2020-21595 & libde265 & Heap Overflow & \multicolumn{1}{c}{\ding{52}} & \multicolumn{1}{c}{\ding{52}} & \multicolumn{1}{c}{\ding{52}} & \multicolumn{1}{c}{/} & \multicolumn{1}{c}{/} & \multicolumn{1}{c}{Build Fail}\\
           
           CVE-2020-21598 & libde265 & Heap Overflow & \multicolumn{1}{c}{\ding{52}} & \multicolumn{1}{c}{\ding{52}} & \multicolumn{1}{c}{\ding{52}} & \multicolumn{1}{c}{/} & \multicolumn{1}{c}{/} & \multicolumn{1}{c}{Build Fail}\\
           
           Issue-5551~\cite{mruby_issue_5551} & mruby & Heap Overflow & \multicolumn{1}{c}{\ding{52}} & \multicolumn{1}{c}{\ding{52}} & \multicolumn{1}{c}{\ding{52}} & \multicolumn{1}{c}{/} & \multicolumn{1}{c}{/} & \multicolumn{1}{c}{Build Fail}\\
           
           CVE-2022-0080 & mruby & Heap Overflow & Run Well & \multicolumn{1}{c}{\ding{52}} & \multicolumn{1}{c}{\ding{52}} & \multicolumn{1}{c}{/} & \multicolumn{1}{c}{/} & \multicolumn{1}{c}{Build Fail}\\
           
           CVE-2019-9021 & PHP & Heap Overflow & \multicolumn{1}{c}{\ding{52}} & \multicolumn{1}{c}{\ding{52}} & \multicolumn{1}{c}{\ding{52}} & \multicolumn{1}{c}{/} & \multicolumn{1}{c}{/} & \multicolumn{1}{c}{Build Fail}\\
           
           CVE-2022-31627 & PHP & Heap Overflow & \multicolumn{1}{c}{\ding{52}} & \multicolumn{1}{c}{\ding{52}} & \multicolumn{1}{c}{\ding{52}} & \multicolumn{1}{c}{/} & \multicolumn{1}{c}{/} & \multicolumn{1}{c}{Build Fail}\\
           
           CVE-2021-32281 & gravity & Heap Overflow & \multicolumn{1}{c}{\ding{52}} & \multicolumn{1}{c}{\ding{52}} & \multicolumn{1}{c}{\ding{52}} & \multicolumn{1}{c}{/} & \multicolumn{1}{c}{/} & \multicolumn{1}{c}{Build Fail}\\
        %[deltatags-prop] Error: unhandled ext func that returns pointer:
           CVE-2020-15888 & Lua & Heap Overflow & \multicolumn{1}{c}{\ding{52}} & \multicolumn{1}{c}{\ding{52}} & \multicolumn{1}{c}{\ding{52}} & \multicolumn{1}{c}{/} & \multicolumn{1}{c}{/} & \multicolumn{1}{c}{Build Fail}\\
           
           CVE-2021-26259 & htmldoc & Heap Overflow & \multicolumn{1}{c}{\ding{52}} & \multicolumn{1}{c}{\ding{54}} & \multicolumn{1}{c}{\ding{52}} & \multicolumn{1}{c}{/} & \multicolumn{1}{c}{/} & \multicolumn{1}{c}{Build Fail}\\
           
           CVE-2022-28966 & Wasm3 & Heap Overflow & \multicolumn{1}{c}{\ding{52}} & \multicolumn{1}{c}{\ding{52}} & \multicolumn{1}{c}{\ding{52}} & \multicolumn{1}{c}{/} & \multicolumn{1}{c}{/} &  \multicolumn{1}{c}{Build Fail}\\
       \bottomrule
        \end{tabular}
        \caption{The security evaluation results of CAMP and related tools on real-world vulnerabilities. \ding{52} represents that the corresponding tool successfully detected the memory corruption in the vulnerability. \ding{54} indicates the tool failed to detect the memory corruption that happened. "/" represents the tool does not support protecting the corresponding type of vulnerability. "Run Well" means the application runs well without causing any memory corruption with the PoC input. "Run Fail" represents that the tool failed to run due to compatibility issues. "Build Fail" means the tool failed to compile the targeted application to enforce protection.}
        \vspace{-2ex}
        \label{tab:realworld-applications}
\end{table*}

% Deltatags Error in Delta pointer means during LTO stage, delta pointer cannot correctly handle func that returns pointer and raise an Error from deltatags\-prop.

\section{Evaluation}
\label{sec:eval}
In this section, 
% we evaluate the security and performance of \sys. In the following, 
we first evaluate \sys's effectiveness in detecting heap overflow and use-after-free memory corruption on a standard vulnerability benchmark and a set of real-world vulnerabilities. Then, we show \sys's protection in detail with two case studies on real-world vulnerabilities. After this, we discuss \sys's security capability with a comparison to related works. Finally, we show \sys's performance/memory overhead using SPEC CPU benchmarks and real-world applications, and demonstrate its advantage against tools from the most recent research. All the experiments were conducted on a bare-metal machine configured with Ubuntu 22.04 system,  12th Gen Intel i7-12700 CPU at 4.9 GHz, 32GB RAM, and 1T SSD storage.

\subsection{Security Evaluation}
\label{sec:sec_eval}
% We evaluate \sys's security in terms of detecting heap overflow and use-after-free. Following some recent works~\cite{}, % dangzero and pacmem
% we used Juliet Test Suite and collected some real-world vulnerabilities to evaluate the effectiveness of \sys. 

\paragraph{Juliet Test Suite.} To evaluate the effectiveness of \sys's protection, we conducted experiments using the Juliet Test Suite, following recent works~\cite{dangzero, li2022pacmem}. The Juliet Test Suite includes test programs for various vulnerability types, each with both bad and good tests. The proof-of-concept (PoC) in the bad tests triggers the corresponding vulnerability, while the PoC in the good tests does not.

As \sys focuses on preventing heap memory corruption, we only included heap-related vulnerability types from the Juliet Test Suite. It is important to note that \sys's overflow prevention mechanism is based on the size of the memory object. Therefore, heap overflows that do not exceed the memory boundary are treated as benign because they do not corrupt other memory objects. Following this logic, we used a customized ASAN\footnote{A customized ASAN that rounds up each allocation and includes a large red zone to prevent overflows from affecting adjacent objects and escaping detection. When evaluating tests for overflow, if ASAN does not report any issues, it indicates that the overflow occurs within an object. In such cases, the test can be safely removed from the test suite.} to exclude the bad tests in which the overflow is contained within the memory object. 

Regarding \sys's use-after-free protection, it neutralizes dangling pointers, so instead of reporting an error, dereferencing a dangling pointer will cause the program to abort without a report. To evaluate the test cases in the use-after-free category, we used a gdb script to confirm that the abortion of the program was due to the dereference of neutralized dangling pointers.

The results of the selected tests are presented in Table~\ref{tab:juliet}, which shows the selected vulnerability type, the number of selected tests, the total number of tests, and the number of tests passed. Some of the tests originally categorized as Heap-Based Buffer Overflow do not contain heap overflow, such as cases \code{Heap_Based_Buffer_Overflow__c_src_char_cat_*}. These tests trigger a buffer overflow when copying data from the heap to the stack, causing a stack overflow rather than a heap overflow. These cases were excluded from the selected test cases using the customized ASAN. In all selected tests, \sys passed without producing any false-positives or false-negatives.

% includes dataset and methodology
% two datasets: Juliet dataset and a set of real-world applications
\paragraph{Real-world Applications.}
% Other than Juliet Test Suite, we used a set of real-world vulnerabilities to further evaluate \sys's effectiveness. We included all the real-world vulnerabilities used in FFmalloc~\cite{}. Since FFmalloc only prevents use-after-free, we searched from the CVE database~\cite{} and collect other types of vulnerabilities. Table~\ref{tab:realworld-applications} shows the list of vulnerabilities. All the vulnerabilities are reproducible with either a working PoC or exploit. The final dataset includes xx use-after-free and xx Heap Overflow bugs across xx applications including language interpreter (e.g., PHP, mruby, Lua, gravity, Wasm3), commonly used library (e.g., libmimedir, libtiff, libpng, libjpeg-turbo, libde265, FFmpeg), browser (Chrome), web server (Nginx), commonly used UNIX tools (e.g., yasm, nasm, cflow, vim, sudo, htmldoc.). We used this rich set of applications aiming to evaluate \sys's effectiveness in preventing different heap memory corruption vulnerabilities and its scalability over different real-world applications.
In addition to the Juliet Test Suite, we also evaluated the security of \sys by using a set of real-world vulnerabilities. We included all the real-world vulnerabilities used in~\cite{ffmalloc}. Additionally, we collected other types of vulnerabilities from the CVE database~\cite{cve_details, cve_program, vuln_db}. Table~\ref{tab:realworld-applications} lists the selected vulnerabilities. Our dataset includes 14 use-after-free and 18 heap overflow vulnerabilities across 19 applications, including language interpreters, commonly used libraries, browsers, web servers, and commonly used UNIX tools. Our goal was to evaluate \sys's effectiveness in preventing different heap memory corruption vulnerabilities and its scalability over various real-world applications. As a comparison, we also evaluate related tools, including ASAN/ASAN\code{--}~\cite{asan_junxu}, Memcheck~\cite{memcheck}, DangNull~\cite{dangnull}, MarkUs~\cite{markus}, and Delta Pointer~\cite{kroes2018delta} for their effectiveness in detecting and preventing heap errors.

Table \ref{tab:realworld-applications} presents the security evaluation results on real-world applications. \sys successfully detected and prevented all use-after-free vulnerabilities. In the case of heap overflow vulnerabilities, \sys was able to detect 16 out of 18 and report them. The other two ran well and did not cause any reports or crashes. But upon manual investigation using a debugger, we found that the overflow had occurred, but the memory bounds were not exceeded due to the rounded-up memory allocation of \sys's seglist allocator. As a result, the overflow is mitigated and no memory corruption happened.
We argue that these two cases do not count as false negatives of \sys as the exploitation is prevented.

For tools providing a comparable level of heap protection, Memcheck was able to detect all the heap errors in the dataset. ASAN\code{--} reported all the heap errors except CVE-2021-26259~\cite{cve-2021-26259}. The reason behind this is that ASAN uses red zone to detect heap overflow vulnerabilities. However, CVE-2021-26259 is a non-linear heap overflow that skips the red zone, thus ASAN\code{--}'s detection is defeated. Unlike ASAN\code{--}, \sys was able to detect this case successfully as it detects heap overflow based on the memory boundary, making it impossible for non-linear heap overflows to bypass its protection.
We argue that \sys's protection is stronger and more robust than those two tools. As discussed in prior work~\cite{li2022pacmem}, ASAN and Memcheck's use-after-free protection could be defeated if the attacker reclaims the freed memory that the dangling pointer refers to, thus enabling a possible exploitation against use-after-free vulnerabilities. As a comparison, \sys mitigates the heap error fundamentally by naturalizing all dangling pointers.

For tools offering partial heap protection, DangNull and Delta Pointer showed limited compatibility support. 4 out of 14 use-after-free cases were not able to build with DangNull. Among 10 use-after-free cases that could be successfully compiled, only 4 of them were detected. Others just crashed with the PoC input as if there is no protection. Note that DangNull has a similar use-after-free protection scheme as \sys, but it fails to detect 6 cases that \sys could detect. This is because DangNull only tracks the point-to relation from heap to structured object on the heap, which will miss the detection if the dangling pointer is on stack/global memory, or the use-after-free object has no type information. MarksUs showed better compatibility support but failed to detect 6 out of 14 use-after-free vulnerabilities.
Delta Pointer showed even worse compatibility support, which can only compile 5 out of 17 cases, but all the out-of-bound in compatible cases were detected. However, due to its design weakness, it is not capable of detecting buffer underflow.

We argue that \sys provides a much more comprehensive heap error detection capability when compared to similar tools. Our evaluation demonstrates that \sys outperforms the combination of partial heap protection tools (such as Delta Pointer + DangNull/MarkUS) as well as ASAN\code{--}.
It is worth mentioning that \sys does not miss any bugs that ASAN\code{--} can detect. In addition, \sys outperforms ASAN\code{--} from the following aspects. First, ASAN\code{--}'s detection for use-after-free is fragile, which could be bypassed by reclaiming free memory with new heap allocation. As such, some use-after-free vulnerabilities in allocation-intensive applications could not be detected. We observed such a case when evaluating ASAN\code{--} in 600.perlbench of SPEC CPU2017, where ASAN\code{--} missed the bug but \sys did not. Besides, some use-after-free POCs that accidentally re-occupy the freed memory will not be reported by ASAN\code{--}. We tweaked the original POC for CVE-2015-2787 and CVE-2015-6838 and found that the use-after-free was missed by ASAN\code{--} as the freed memory is reclaimed. Second, ASAN\code{--} utilizes a red zone mechanism to flag out-of-bound access. In the case of non-linear overflow, the overflow may jump over the red zone and thus fail overflow detection. Although it is unknown how frequently non-linear overflow may happen in the real world, we argue that the missed detection of such bugs will cause serious security issues. For example, CVE-2021-26259 is a non-linear overflow that would lead to code execution. 
% Going beyond Failing to pinpoint non-linear overflow could also lead to additional effort to identify the root cause and fix the issue. 
% Assuming that the non-linear overflow corrupts data in the adjacent object and causes memory corruption somewhere else, the developers will need to figure out the root cause of unrelated bug reports.

\begin{table*}[t]
% \vspace{-20pt}
    \tabcolsep=3pt
    \centering
    \begin{tabular}{lccccc}
        \toprule
        \multicolumn{1}{c}{\multirow{2}{*}{\textbf{Benchmark}}} & \multicolumn{5}{c}{\textbf{Time and Memory Overhead}}  \\
        \cline{2-6} \rule{0pt}{11pt}
        & \textbf{CAMP}     & \textbf{ASAN$--$}  & \textbf{ASAN}     & \textbf{ESAN}      & \textbf{Memcheck} \\        
        \midrule

        600.perlbench\_s & 237.95\% / 2241.12\% & 76.95\% / 366.92\% & 143.59\% / 358.20\% & 644.00\% / 4.15\% & 3496.46\% / 138.97\% \\
        602.gcc\_s & 78.56\% / 135.52\% & 83.61\% / 63.42\% & 99.47\% / 62.77\% & - & 2888.13\% / 30.42\% \\
        605.mcf\_s & 14.62\% / 31.55\% & 24.45\% / 3.61\% & 27.88\% / 3.61\% & 109.33\% / -4.24\% & 601.05\% / 22.68\% \\
        623.xalancbmk\_s & 138.94\% / 1220.66\% & 107.86\% / 428.07\% & 109.41\% / 433.51\% & 81.67\% / 8.60\% & 4962.60\% / 98.81\% \\
        625.x264\_s & 75.07\% / 12.68\% & 62.26\% / 13.52\% & 75.92\% / 13.26\% & 90.94\% / -3.55\% & 2070.57\% / 56.96\% \\
        631.deepsjeng\_s & 1.58\% / 0.00\% & 44.23\% / -0.23\% & 64.08\% / -0.23\% & 18.85\% / -0.25\% & 3251.34\% / 25.34\% \\
        641.leela\_s & 3.02\% / 514.19\% & 13.97\% / 2832.83\% & 17.33\% / 2833.72\% & 6.65\% / -17.52\% & 4163.69\% / 262.82\% \\
        657.xz\_s & 7.79\% / 0.00\% & 17.45\% / 2.98\% & 13.40\% / 2.98\% & 14.61\% / -0.70\% & 718.87\% / 24.45\% \\
        619.lbm\_s & 1.34\% / 0.01\% & 37.32\% / 5.94\% & 29.38\% / 5.94\% & 34.14\% / -0.36\% & 2907.53\% / 25.98\% \\
        638.imagick\_s & 45.47\% / 0.07\% & 17.23\% / 4.46\% & 28.56\% / 4.47\% & 21.70\% / -2.00\% & 4452.66\% / 22.93\% \\
        644.nab\_s & 62.55\% / 26.13\% & 35.18\% / 67.52\% & 35.14\% / 66.63\% & 1988.66\% / -1.34\% & 3722.35\% / 31.80\% \\
        \midrule
        Geomean & 21.27\% / 127.47\% & 38.27\% / 104.72\% & 44.78\% / 104.35\% & 65.31\% / -1.94\% & 2546.88\% / 56.49\% \\

        \bottomrule
    \end{tabular}
    \caption{The relative time and memory overhead of CAMP, ASAN~\code{--}, ASAN, ESAN, and Memcheck on SPEC CPU2017. "-" indicates the tool failed to run the corresponding benchmark.}
    \label{tab:overhead-cpu2017}
\end{table*}

\begin{table*}[t]
    \centering    
    \begin{tabular}{ccccccccc}
        \toprule
        \textbf{Benchmark}         & \textbf{Metric} &
        \textbf{CAMP} &
        \textbf{LowFat} &
        \textbf{Delta Pointer} &
        \textbf{DangNull} &
        \textbf{FreeGuard} &
        \textbf{MarkUs} &
        \textbf{FFMalloc} \\
        \midrule
        \multirow{2}{*}{SPEC CPU2006} 
        & Time
        & 54.92\% & 160.62\% & 37.39\% & 39.99\% & 10.40\% & 15.84\% & 9.50\% \\ \cline{2-2} \rule{0pt}{11pt}                         
        & Mem          
        & 237.67\% & 38.60\% & 0.01\% & 158.52\% & 70.89\% & 2.96\% & 27.57\% \\ \midrule
        \multirow{2}{*}{SPEC CPU2017} 
        & Time            
        & 21.27\% & 96.96\% & - & 28.61\% & 8.23\% & 11.90\% & 10.94\% \\ \cline{2-2} \rule{0pt}{11pt}
        & Mem          
        & 127.47\% & 54.35\% & - & 314.13\% & 29.23\% & 32.35\% &  62.00\% \\ 
        \bottomrule
    \end{tabular}
    \caption{The relative time and memory overhead of CAMP, LowFat, Delta Pointer, DangNull, FreeGuard, MarkUs, and FFmalloc on SPEC CPU2006 and SPEC CPU2017.}
    \vspace{-2ex}
    \label{tab:camp_uaf_oob}
\end{table*}

% \begin{table*}[t]
%     \centering    
%     \begin{tabular}{cccccccc}
%         \toprule
%         \textbf{Benchmark}         &
%         \textbf{CAMP} &
%         \textbf{LowFat} &
%         \textbf{Delta Pointer} &
%         \textbf{DangNull} &
%         \textbf{FreeGuard} &
%         \textbf{MarkUs} &
%         \textbf{FFMalloc} \\
%         \midrule
%         \multirow{1}{*}{SPEC CPU2006} 
%         & 54.92\% & 160.62\% & 37.39\% & 39.99\% & 10.40\% & 15.84\% & 9.50\% \\   
%         \multirow{1}{*}{SPEC CPU2017}            
%         & 21.27\% & 96.96\% & - & 28.61\% & 8.23\% & 11.90\% & 10.94\% \\  
%         \bottomrule
%     \end{tabular}
%     \caption{Time overhead of CAMP, LowFat, Delta Pointer, DangNull, FreeGuard, MarkUs and FFmalloc on SPEC CPU2006 and SPEC CPU2017.}
%     \label{tab:camp_uaf_oob}
% \end{table*}

\begin{figure*}[t]
\vspace{-10pt}
\includegraphics[width=\textwidth]{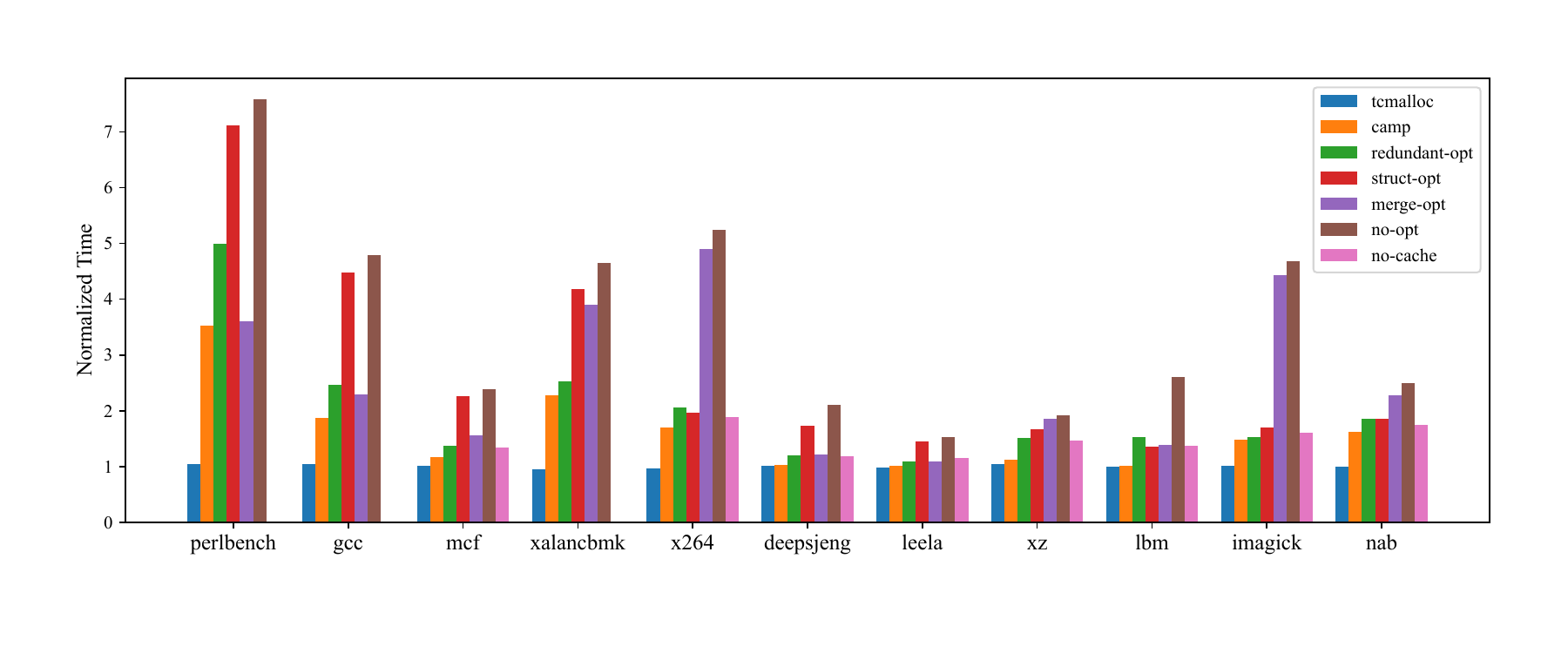}
\vspace{-40pt}
\caption{Evaluation result of CAMP breakdown on SPEC CPU2017. From left to right, the bars show the normalized time of tcmalloc replacement, CAMP, CAMP with each optimization disabled, and CAMP without optimization.}
\vspace{-2ex}
\label{fig:opt}
\end{figure*}

\subsection{Performance Evaluation}

In the following, we evaluate \sys's performance on the two SPEC CPU benchmarks with the comparison with related tools.
Then, we evaluate the effectiveness of each design component of \sys, including the compiler optimization, and the customized allocator.
Finally, we evaluate \sys on the real-world applications.

\paragraph{SPEC CPU Benchmark.} We compared the performance of \sys with related tools including ASAN~\cite{address_sanitizer}, ASAN~\code{--}\cite{asan_junxu}, ESAN~\cite{duck2018effectivesan}, Softbound+CETS~\cite{softbound}, and Memcheck~\cite{memcheck}. All programs in the benchmark suite were compiled using default configurations and used reference input. Besides, all tools were configured to ignore detected errors to avoid termination.
Note that, to ensure a fair comparison, we only enabled ASAN~\code{--} and ASAN's heap error detector.
Unfortunately, ASAN and Memcheck were unable to compile and run the \texttt{omnetpp} and \texttt{dealII} programs, and therefore, they were excluded from the evaluation. Softbound+CETS showed limited support, failing to compile all programs in SPEC CPU2017 and only supporting 7 programs in SPEC CPU2006, so we only compare with it in SPEC CPU2006. Our evaluation of the tools did not take PACMem~\cite{li2022pacmem} into account, as it requires specialized hardware (ARM PA) for the detection of heap memory errors. For baseline evaluation, we used tcmalloc as the default allocator for better comparison as \sys's allocator is based on tcmalloc. For each tool, we ran the benchmark 10 times and reported the average result to minimize the randomness.

The evaluation results in terms of time overhead and memory overhead are presented in Table~\ref{tab:overhead-cpu2017} for the SPEC CPU2017 benchmark suite. Each row in the table represents a specific application benchmark, with the benchmark name listed in the first column and the subsequent columns display the relative time and memory measured compared to a baseline. Following the most recent works~\cite{li2022pacmem, dangzero}, we utilized geometric mean value to represent the average overhead of each tool.
\sys exhibits the best runtime speed compared to other tools, with an average overhead of 21.27\%, while ASAN~\code{--}, ASAN, and ESAN have overhead rates of 38.27\%, 44.78\%, and 65.31\%, respectively. Memcheck has the worst runtime performance, which introduces 2546.88\% overhead compared to the baseline. In terms of memory overhead, \sys has a higher rate of 127.47\% compared to roughly 104\% for ASAN and ASAN~\code{--}. ESAN has the best memory overhead performance, with -1.94\% which is mainly cased by the difference of allocators. As we discussed in Section~\ref{sec:design}, \sys tracks different forms of object point-to relation, which is more comprehensive and could obtain better security guarantee as we showed in Section~\ref{sec:sec_eval}. The cost of this design is that the performance overhead will be higher on allocation-intensive programs, such as \code{600.perlbench_s} and \code{623.xalancbmk_s}. We found that \sys recorded a number of escape tracks, resulting in notable time and memory overhead.
% Specifically, \code{600.perlbench_s} exhibited a time overhead of 237.95\% and a memory overhead of 2241.12\%. Similarly, \code{623.xalancbmk_s} experienced a time overhead of 138.94\% and a memory overhead of 1220.66\%. 
The average time and memory overhead of \sys without those two cases will be reduced to 13.21\% and 44.24\%, respectively. For case \code{638.imagick_s} and \code{644.nab_s}, where \sys's time overhead is higher than ASAN\code{--}, we observed the proposed optimizations are less effective in those two cases. Specifically, the type-based range checks removing and redundant checks removing are less effective on them. For the remaining cases in the benchmark, \sys's time overhead is minimal with an even lower geomean value of 8.86\%.

As for SPEC CPU2006, the evaluation results could be found in Table~\ref{tab:time-overhead-cpu2006} and Table~\ref{tab:memory-overhead-cpu2006}, we put them in Appendix due to the limited space. \sys still outperforms all other tools in terms of runtime speed. Specifically, it introduces an overhead of 54.92\%, while ASAN~\code{--}, ASAN, ESAN, SoftBound+CETS, and Memcheck have the overhead of 56.77\%, 67.02\%, 123.08\%, 319.75\%, and 1990.02\%, respectively. As we discussed before, \sys performs worse in allocation-intensive programs. As such, for cases like \code{400.perlbench}, \code{482.sphinx3}, \code{453.povray}, \code{453.povray}, \code{473.astar}, \code{483.xalancbmk}, \sys traces a large number of runtime point-to relations and costs more memory on maintaining them, which eventually slows down the overall speed. The time and memory overhead without those cases will be reduced to 34.52\%, and 39.3\%, respectively. Since the SPEC CPU2006 contains more allocation-intensive cases than SPEC CPU2017, \sys reports higher overhead on SPEC CPU2006.

We further compare \sys with other tools that offer partial memory protection, following the same setup of the previous evaluation on the SPEC CPU Benchmarks, we present the results in Table~\ref{tab:camp_uaf_oob}. LowFat, which only provides out-of-bounds (OOB) protection, exhibits slower performance than \sys, with time overheads of 160.62\% and 96.96\% on the two SPEC benchmarks, respectively. Delta Pointer, which could only run SPEC CPU2006, incurs 37.39\% time overhead. In terms of use-after-free (UAF) protections, FreeGuard, MarkUs, and FFMalloc introduce approximately 10\% overhead on the benchmarks. DangNull exhibits 28.61\% overhead on SPEC CPU 2017, which is slightly high than \sys. It is noted that DangNull only includes a similar but weaker use-after-free protection scheme, highlighting the fast runtime speed provided by the \sys allocator while offering a stronger security guarantee.
We argue that \sys is the optimal solution for achieving comprehensive heap protection in comparison to the other tools currently available. Firstly, as opposed to tools that only offer partial heap protection, \sys presents a more holistic solution, with the ability to detect both spatial and temporal heap errors.
Moreover, even when considering the combination of the faster Out-of-Bounds protection tool (e.g., Delta Pointer) and Use-After-Free protection tool (e.g., FFMalloc), the overhead incurred by \sys is remarkably similar to their collective overhead (54.92\% vs. 46.89\%). Despite this, \sys outperforms this combination in several ways. On the one hand, due to design limitations, Delta Pointer is only capable of detecting overflow errors, leaving underflow errors undetected, a challenge that \sys effectively addresses. On the other hand, Delta Pointer's support is restricted to a 32-bit address space, allowing for a maximum of only 4GB of memory. In parallel, FFMalloc may demand a considerable amount of physical memory if the program continues to persist~\cite{dangzero}. Hence, their combined use could rapidly exhaust the available 4GB of memory, resulting in incompatibility issues.

\begin{table}[t]
    \centering
    \begin{tabular}{lccc}
        \toprule
        \multirow{2}*{\textbf{Benchmark}} &        
        \multicolumn{2}{c}{\textbf{Time ($s$)}} &
        \multirow{2}*{\textbf{Overhead}} \\
       \cline{2-3} \rule{0pt}{11pt} &
        \textbf{Native} & 
        \textbf{CAMP} \\
        \midrule
        cfrac          & 2.91 & 3.82 & 31.27\%  \\
        espresso       & 3.62 & 3.62 & 0.00\%   \\
        barnes         & 1.35 & 1.34 & -0.74\%  \\
        redis          & 2.66 & 2.68 & 0.75\%   \\
        leanN          & 25.35 & 26.18 & 3.27\%   \\
        alloc-test1    & 2.99 & 3.06 & 2.34\%   \\
        alloc-testN    & 2.91 & 3.42 & 17.53\%  \\
        sh6benchN      & 2.41 & 2.39 & -0.83\%  \\
        sh8benchN      & 5.79 & 8.3 & 43.35\%  \\
        xmalloc-testN  & 2.664 & 2.306 & -13.44\% \\
        cache-scratchN & 0.43 & 0.44 & 2.33\%  \\
        \midrule
        Geomean & - & - & 9.79\% \\
        \bottomrule
    \end{tabular}
    \caption{Time Overhead on mimalloc-bench. Native represents using the default allocator -- ptmalloc, CAMP means using its customized seglist allocator based on tcmalloc.}
    \vspace{-2ex}
    \label{tab:mibench}
\end{table}

\paragraph{Component Evaluation.} The configuration of \sys contains various components, including compiler optimization and the integration of a seglist allocator. To determine the impact of each component on performance, we conducted evaluations using different setups of \sys.

To analyze the impact of \sys's customized allocator, we conducted an evaluation using mimalloc-bench to compare the allocator differences. The evaluation results are presented in Table~\ref{tab:mibench}. The "Native" column represents the performance of the system's native allocator (ptmalloc), while the "\sys" column indicates the results obtained with \sys's customized allocator. The two allocators exhibited distinct behavior across different test cases. For instance, in the case of \code{cfrac}, the Native allocator was 31.27\% faster than \sys, whereas in the \code{xmalloc-testN} case, it was 13.44\% slower. Overall, \sys's allocator demonstrated a 9.79\% slowdown compared to the native allocator.

To evaluate the effectiveness of compiler optimization, we disabled each optimization one by one in different configurations of \sys, represented as \code{struct-opt}, \code{redundant-opt}, \code{merge-opt} and \code{no-opt} in Figure~\ref{fig:opt}. To gain insight into the role of the allocator cache design, we measured the performance of \sys allocator cache disabled. Finally, to assess the impact of the seglist allocator, we compared the results to a baseline using tcmalloc. These setups and their results on SPEC CPU2017 are presented in Figure~\ref{fig:opt}.

Our evaluation results confirm that the seglist allocator offers minimal benefit. The tcmalloc baseline, compared to default ptmalloc, only improves average speed by 2.26\%. We then took a further look at the contribution of allocator cache design, we found three programs (\code{perlbench}, \code{gcc}, \code{xalancbmk}) experienced memory exhaustion and were unable to complete the test, so they were excluded from Figure~\ref{fig:opt}. The remaining programs resulted in an average overhead of 40.34\%, nearly double the overhead with cache enabled (20.94\%). Our investigation revealed that these three programs made many repeated point-to relationships, leading to a high consumption of memory for metadata maintenance. 
Specifically, all these programs contain a language interpreter that builds ASTs during input parsing, which creates connections that require \sys to maintain point-to records. In \code{gcc}, as the compiler optimization progressed, more nodes are connected, leading to more point-to records being constructed, and eventually, the program exhausted all the memory available. In \code{perl}, after parsing the AST, the execution of the AST utilized the heap as a stack, leading to repeated references that further constructed more pointer escapes. These cases demonstrate the effectiveness of the cache design. With it, repeated point-to relations could be saved, thus reducing both memory and time overhead.

In addition, we found that the proposed compiler optimization significantly reduced the performance overhead. The \sys without compiler optimization imposed an overhead of 204.09\%, while the default \sys had an overhead of 20.94\%. Among the tested programs, \code{imagick} saw the best optimization results, with its overhead reduced from 368.18\% to 48.47\%. Upon analyzing the breakdown of each optimization, we found that the greatest impact came from the structure optimization. If this optimization is disabled, the overhead increases from 20.94\% to 113.39\%. While disabling redundant optimization, and merging runtime call optimization introduced 64.20\%, and 95.73\%, respectively.

\begin{table}[t]
    % \vspace{-20pt}
    \tabcolsep=4.5pt
    \centering
    \begin{tabular}{ccccccc}
       \toprule
        \multirow{2}*{\textbf{System}} &
        \multirow{1}*{\textbf{Output}} &
        \multicolumn{5}{c}{\textbf{Latency ($\mu s$)}} \\
        \cline{3-7} \rule{0pt}{11pt} & ($req/s$)
        & \textbf{Average} & \textbf{50\%} & \textbf{75\%} 
        & \textbf{90\%} & \textbf{99\%} \\
       \midrule
        Native & 150,368 & 643.23 & 625 & 635 & 649 & 910 \\
        CAMP & 108,322 & 880 & 850 & 870 & 910 & 1070 \\ 
        ASAN\code{--} & 103,688 & 930 & 880 & 900 & 960 & 1860 \\ 
        ASAN & 97,095 & 970 & 900 & 930 & 1040 & 1910 \\ \midrule
    \end{tabular}
    \caption{CAMP, ASAN, and ASAN\code{--}'s output and latency evaluation results on Nginx. In the Latency column, the "Average" represents the average latency of the requested connection, the others show the latency distribution. 
    % On average, CAMP imposes 27.96\% overhead on Nginx's request output and 36.81\% more time on latency. Whereas ASAN introduces 35.43\% and 50.80\% overhead on output and latency, respectively.
    }
    \vspace{-2ex}
    \label{tab:nginx_eval}
\end{table}

\paragraph{Nginx.} To evaluate the performance of \sys on a large-scale, real-world application, we conducted experiments on Nginx v1.22.1 using the wrk v4.2.0 benchmarking tool. For these experiments, we configured the tool with 8 threads, 100 connections, and a test duration of 60 seconds. To ensure consistency, we repeated the test 30 times and recorded the average results. The findings are presented in Table~\ref{tab:nginx_eval}. On average, CAMP introduces a 27.96\% overhead on Nginx's request output, In terms of latency, \sys adds 36.81\% more time. The results reflect \sys's efficiency on real-world applications with mild overhead. As a comparison,  ASAN and ASAN\code{--} incur a 35.43\% and 31.04\% overhead on request output and have a latency overhead of 50.80\% and 44.58\%, respectively.

\paragraph{Chromium.} In addition to Nginx, we also evaluated the performance of \sys on Chromium. We used three popular browser benchmarks: Kraken, SunSpider, and Lite Brite. Furthermore, we measured the loading time for websites, as this metric is highly relevant to end users' browsing experience. To measure loading time, we utilized a browser extension and recorded the average loading time for the 8 most popular websites according to the Top Websites Ranking~\cite{top_website}. Each benchmark experiment was repeated 30 times, and the mean value was calculated to mitigate any random variations. The evaluation results are presented in Table~\ref{tab:eval_chrome}, with the average overhead of CAMP represented by the geometric mean.

We observed that CAMP introduced a 67.14\% overhead on the three browser benchmarks. However, the loading time for web pages increased by only 28.59\% on average. It is important to note that these benchmarks focus on specific components of the browser, which may not fully reflect the overall performance. In contrast, loading web pages involves JavaScript engine execution, DOM processing, and other factors, thus it is more representative of real-world browsing. In this regard, we argue that the overhead introduced by \sys to Chromium is minimal.

\begin{table}[t]
    % \vspace{-20pt}
    \tabcolsep=4.5pt
    \centering
    \begin{tabular}{cccc}
       \toprule
        \multirow{2}*{\textbf{Benchmark}} &        
        \multicolumn{2}{c}{\textbf{Time ($ms$)}} &
        \multirow{2}*{\textbf{Overhead}} \\
        \cline{2-3} \rule{0pt}{11pt} &
        \textbf{Native} & 
        \textbf{CAMP} \\
       \midrule
        kraken & 1069 & 1722 & 61.09\% \\
        sunspider & 521 & 813 & 56.05\% \\
        Lite Brite & 2930 & 5520 & 88.40\% \\
       \midrule
       Geomean & - & - 
       & 67.14\%  \\
       \toprule
        google.com & 1101 & 1427 & 29.61\% \\
        facebook.com  & 831  & 1199 & 44.28\% \\
        amazon.com    & 2298 & 3120 & 35.77\% \\
        openai.com    & 1444 & 1791 & 24.03\% \\
        twitter.com   & 1479 & 1708 & 15.48\% \\
        gmail.com     & 1691 & 2032 & 20.17\% \\
        youtube.com   & 2143 & 2628 & 22.63\% \\
        wikipedia.org & 984  & 1535 & 56.00\% \\
       \midrule
       Geomean & - & - 
       & 28.59\% \\
       \bottomrule
    \end{tabular}
    \caption{CAMP's performance evaluation results on the Chromium browser. In the Benchmark column, kraken, sunspider and Lite Brite are three browser benchmarks, whereas the following are websites used to measure the loading time of the browser.}
    \vspace{-2ex}
    \label{tab:eval_chrome}
\end{table}

\section{Discussion}
\label{sec:dis}

\paragraph{False Positive and False Negative.} As a pointer-based protection approach, \sys shares similar weaknesses with prior works~\cite{softbound, low_fat, in_fat_pointer}. First, C/C++ allows the use of out-of-bound pointers as the memory boundary, which may cause \sys to generate false positives. To mitigate this, we reserve additional memory space allocated for each allocation so that those boundary pointers will be still in bound, as is discussed in Section~\ref{sec:impl}. This approach of changing the memory layout effectively mitigated this issue, thus we did not find any false positives in our evaluation. We consider this to be more of a compatibility issue that should be resolved at the source code level. This not only eliminates the potential for false positives, but also strengthens the security of the code.

Second, in C/C++, integers may be used for pointer arithmetic instead of actual pointers. This poses a challenge for \sys as it may result in out-of-bounds pointer accesses that go undetected. To address this issue, we enforce a policy within the compiler that prohibits the casting of integers to pointers. Thus, the generation of out-of-bounds pointers from integers can be prevented. In cases where developers have to perform such casts, they could explicitly indicate their intention using compiler attributes to disable the policy locally.

\paragraph{Preventing In-bound Overflow.} The design of \sys's overflow detection is based on memory boundaries, meaning that it only identifies overflows that cross the boundaries as violations. Therefore, \sys's design shares a similar weakness with prior works~\cite{softbound, prober, in_fat_pointer}, as it cannot be used to prevent in-bound heap overflows.  However, we found that, by leveraging a proper implementation, \sys could mitigate some in-bound overflow. Specifically, if pointer arithmetic is performed on an array of a structure, we can use the array size to validate the pointer and ensure that the result pointer stays within the array of the structure, as such it mitigates the in-bound overflow with the type information.
However, we do not claim this as \sys's capability of preventing in-bound overflow. Because the type information is not always available, besides, the array may be dynamic thus \sys has no clue how to validate it to make sure no overflow is inside the structure. Therefore, we consider the prevention of in-bound overflows to be a future research direction.

\paragraph{Protecting Shared Library.} As \sys uses the compiler for instrumentation, it naturally supports the protection of any programs that the compiler can build, including shared libraries. To achieve this protection, users must re-compile the shared library using the \sys compiler. By doing so, the library will be instrumented with function calls to the \sys runtime, ensuring that protection is enforced.

\paragraph{Supporting Multi-threaded Programs.} \sys is compatible with multi-threaded programs. Data race could be avoided by employing locks for the metadata in each span. In addition, we evaluated several multi-threaded programs (e.g., PHP, mruby, Lua, FFmpeg, Chrome, Nginx) in Section~\ref{sec:eval} and did not observe any incompatibility -- \sys successfully detected the bug in those programs without reporting any false positive.

% \paragraph{CAMP's Protection Approach}

% \paragraph{CAMP's Utility} \fixme{discuss whether it is a sanitizer or protection}
\section{Related Work}
\label{sec:related}
% \zip{rewrite this section to include more state of the art, cheri hardware support of protection, include OSCAR, redfat, Delta pointers}
% In this section, we summarize related works and highlight the differences between them.
% safe allocator
% memory error detector
% pointer invalidation
% hardware assisted protection.

\paragraph{Safe allocator.} To combating heap memory corruption attacks~\cite{wang2021maze, heelan2018automatic, lin2022dirtycred, yun2020automatic, elastic_object}, various safe heap allocators~\cite{ffmalloc, dangzero, dang2017oscar, mimalloc, Slimguard, scudo_allocator} are proposed. Specifically, FFmalloc~\cite{ffmalloc} proposes a one-time memory allocation. DieHarder~\cite{dieharder} randomizes heap address space to unstabilize heap exploitation. Oscar~\cite{dang2017oscar} prevents temporal memory error with shadow memory allocated for each heap object, thus detecting dangling pointer access. Markus~\cite{markus} employs a strategy that quarantines freed memory to eliminate any lingering dangling pointers. DangZero~\cite{dangzero} interacts with the kernel page table to implement an alias-based UAF detection mechanism, ensuring the virtual memory is never reused. FreeGuard~\cite{freeguard} improves performance with freelist and optimizes shadow memory scheme. \sys differs from those works as it does not exclusively rely on allocators to enable protection.

\paragraph{Pointer invalidation.} Several works~\cite{softbound, cets, dangnull, undangle, prober, kroes2018delta, redfat} detect memory errors through pointer invalidation. For instance, CETS~\cite{cets} employs a lock-and-key identifier-based approach to track separate metadata for each pointer to detect dangling pointers. Undangle~\cite{undangle} uses dynamic taint tracking to identify and eliminate unsafe dangling pointers. DangNull~\cite{dangnull} nullifies pointers when their associated objects are freed. In regards to spatial safety, Redfat~\cite{redfat} combines redzone and low-fat pointers to detect buffer overflow. Delta Pointer~\cite{kroes2018delta} introduces a pointer tag to invalidate overflow pointers, thereby preventing errors.  Softbound~\cite{softbound} utilizes shadow memory to track memory bounds and employs run-time checks for efficient overflow detection. \sys achieves full heap protection with pointer validation, however, unlike the aforementioned works, \sys cooperates with the compiler and the allocator to optimize pointer invalidation.

\paragraph{Memory Sanitizer.} Memory sanitizers typically offer full heap error detection. ASan~\cite{address_sanitizer} employs shadow memory and redzones for detecting temporal and spatial errors. To reduce its overhead, ASan\code{--}~\cite{asan_junxu} and SANRAZOR~\cite{zhang2021sanrazor} propose several compiler optimizations to reduce instrumented checks on memory access, without security compromise. FuZZan~\cite{jeon2020fuzzan} presents new metadata structures to decrease memory management overhead. CUP~\cite{burow2018cup} proposes a hybrid metadata scheme that supports all program data including globals, heap, and stack. EffectiveSan~\cite{duck2018effectivesan} presents a dynamic type system to detect memory errors, but it has some limitations on detecting temporal errors. Memcheck~\cite{memcheck}, part of Valgrind~\cite{valgrind}, detects full memory errors. \sys differentiates itself from those approaches by its protection scheme, thereby offering superior speed.

\paragraph{Hardware-assisted protection.} There are also a bunch of works~\cite{farkhani2021ptauth, saileshwar2022heapcheck, low_fat, kim2020hardware, in_fat_pointer, woodruff2014cheri, zhang2019bogo, li2022pacmem} that leverage hardware to enforce memory safety. Specifically, Low-Fat~\cite{low_fat} extends the pointer representation with base and bounds information so that the runtime or hardware can prevent spatial safety violations. In-Fat~\cite{in_fat_pointer} enhances the hardware-assisted tagged-pointer scheme, employing three complementary object metadata schemes to decrease the number of pointer tag bits required.
Several works, including PtAuth~\cite{farkhani2021ptauth}, AOS~\cite{kim2020hardware}, and PACMem~\cite{li2022pacmem}, utilize the Pointer Authentication Code (PAC) feature of ARM to identify memory errors.
% HeapCheck~\cite{saileshwar2022heapcheck} uses spare pointer bits in 64-bit systems to index a bounds table, facilitating detection of memory errors. Storing bounds info in an 8 KB on-chip SRAM cache, it introduces minimal overhead.
% BOGO~\cite{zhang2019bogo} utilizes Intel MPX to ensure spatial and temporal memory safety. CHEx86~\cite{sharifi2020chex86} presents an innovative process architecture aimed at preventing memory errors by instrumenting the code at the microcode level. Unlike those works, \sys has no requirement for additional hardware support.
HeapCheck~\cite{saileshwar2022heapcheck} leverages pointer bits in 64-bit systems for a bounds table, aiding in memory error detection with an 8 KB on-chip SRAM cache. BOGO~\cite{zhang2019bogo} employs Intel MPX for memory safety, while CHEx86~\cite{sharifi2020chex86} innovatively targets memory errors via microcode-level code instrumentation. In contrast, \sys doesn't need extra hardware support.
\section{Conclusion}
\label{sec:conclusion}

% In conclusion, mitigating memory corruption on the heap is a complex task that requires a balance between security and performance. Existing techniques to address heap memory corruption either provide limited protection or introduce significant runtime overhead, making their adoption in real-world products challenging. Our approach, \sys, combines a customized allocator with compiler-based instrumentation and optimization techniques to provide effective and efficient protection against heap memory corruption. Our evaluation shows that \sys can reduce the runtime overhead to as low as 4\% in a real-world application. \sys presents a promising solution for safeguarding programs against heap memory corruption.

% In conclusion, 
Mitigating memory corruption on the heap is a complex task.
Existing techniques to address heap memory corruption either provide limited protection or introduce significant runtime overhead, making their adoption in real-world products challenging. By leveraging a carefully designed code instrumentation and a customized allocator, \sys provides comprehensive protection against heap memory corruption. The instrumentation imposes some runtime overhead, but we demonstrate that this overhead can be significantly reduced through a series of optimization strategies that eliminate and consolidate unnecessary instrumentations. Our evaluation of \sys using a large-scale real-world application and SPEC CPU Benchmarks shows that the performance impact is significantly reduced. The low overhead, combined with \sys' ability to effectively detect and prevent heap memory corruption, makes it a promising solution for safeguarding programs against heap memory corruption.

\section*{Acknowledgement}
We thank our shepherd and other anonymous reviewers for
their insightful feedback. 
This work was supported by grants
from the Defense Advanced Research Projects Agency (DARPA) under contract No. N6600122C4026,
the Office of Naval Research (ONR) under Grant No. N00014-20-1-2008,
the U.S. National Science Foundation (NSF) via award CCF-2119069, CNS-2211508, CNS-2211315, CNS-1763743, CCF-2028851, CCF-2107042, and CCF-1908488,
and the U.S. Department of Energy (DOE) via project 17-SC-20-SC and DESC0022268.
Any opinions, findings, conclusions, or recommendations expressed in this material are those of the authors and
do not necessarily reflect the views of the funding agency.
\newpage
{
\footnotesize
\bibliographystyle{IEEEtran}
\bibliography{ref}

\begin{thebibliography}{10}
\providecommand{\url}[1]{#1}
\csname url@samestyle\endcsname
\providecommand{\newblock}{\relax}
\providecommand{\bibinfo}[2]{#2}
\providecommand{\BIBentrySTDinterwordspacing}{\spaceskip=0pt\relax}
\providecommand{\BIBentryALTinterwordstretchfactor}{4}
\providecommand{\BIBentryALTinterwordspacing}{\spaceskip=\fontdimen2\font plus
\BIBentryALTinterwordstretchfactor\fontdimen3\font minus \fontdimen4\font\relax}
\providecommand{\BIBforeignlanguage}[2]{{%
\expandafter\ifx\csname l@#1\endcsname\relax
\typeout{** WARNING: IEEEtran.bst: No hyphenation pattern has been}%
\typeout{** loaded for the language `#1'. Using the pattern for}%
\typeout{** the default language instead.}%
\else
\language=\csname l@#1\endcsname
\fi
#2}}
\providecommand{\BIBdecl}{\relax}
\BIBdecl

\bibitem{msr_vuln_stat}
M.~Miller, ``{A snapshot of vulnerability root cause trends for Micrsoft Remote Code Execution (RCE) CVEs, 2006 through 2017},'' \url{https:// twitter.com/epakskape/status/984481101937651713}.

\bibitem{p0_vuln_stat}
G.~P. Zero, ``{0day "In the Wild"},'' \url{https://docs.google.com/spreadsheets/d/1lkNJ0uQwbeC1ZTRrxdtuPLCIl7mlUreoKfSIgajnSyY/edit#gid=0}.

\bibitem{2023kernelcve}
``{Linux Kernel CVE Changes},'' \url{https://www.linuxkernelcves.com/}.

\bibitem{freeguard}
S.~Silvestro, H.~Liu, C.~Crosser, Z.~Lin, and T.~Liu, ``Freeguard: A faster secure heap allocator,'' in \emph{Proceedings of the 2017 ACM SIGSAC Conference on Computer and Communications Security}, 2017.

\bibitem{lin2009polymorphing}
Z.~Lin, R.~D. Riley, and D.~Xu, ``Polymorphing software by randomizing data structure layout,'' in \emph{Detection of Intrusions and Malware, and Vulnerability Assessment: 6th International Conference, DIMVA}.\hskip 1em plus 0.5em minus 0.4em\relax Springer, 2009.

\bibitem{dieharder}
G.~Novark and E.~D. Berger, ``Dieharder: securing the heap,'' in \emph{Proceedings of the 17th ACM conference on Computer and communications security}, 2010.

\bibitem{wang2021maze}
Y.~Wang, C.~Zhang, Z.~Zhao, B.~Zhang, X.~Gong, and W.~Zou, ``$\{$MAZE$\}$: Towards automated heap feng shui,'' in \emph{30th USENIX Security Symposium (USENIX Security 21)}, 2021, pp. 1647--1664.

\bibitem{heelan2018automatic}
S.~Heelan, T.~Melham, and D.~Kroening, ``Automatic heap layout manipulation for exploitation,'' in \emph{27th USENIX Security Symposium (USENIX Security 18)}, 2018, pp. 763--779.

\bibitem{lin2022dirtycred}
Z.~Lin, Y.~Wu, and X.~Xing, ``Dirtycred: Escalating privilege in linux kernel,'' in \emph{Proceedings of the 2022 ACM SIGSAC Conference on Computer and Communications Security}, 2022.

\bibitem{yun2020automatic}
I.~Yun, D.~Kapil, and T.~Kim, ``Automatic techniques to systematically discover new heap exploitation primitives,'' in \emph{29th USENIX Security Symposium (USENIX Security 20)}, 2020, pp. 1111--1128.

\bibitem{elastic_object}
Y.~Chen, Z.~Lin, and X.~Xing, ``A systematic study of elastic objects in kernel exploitation,'' in \emph{Proceedings of the 2020 ACM SIGSAC Conference on Computer and Communications Security}, 2020.

\bibitem{memcheck}
N.~N. Julian~Seward, ``{{Memcheck: a memory error detector}},'' \url{https://valgrind.org/docs/manual/mc-manual.html}.

\bibitem{address_sanitizer}
K.~Serebryany, D.~Bruening, A.~Potapenko, and D.~Vyukov, ``Addresssanitizer: A fast address sanity checker,'' in \emph{Proceedings of the 2012 USENIX Conference on Annual Technical Conference}, 2012.

\bibitem{dangnull}
B.~Lee, C.~Song, Y.~Jang, T.~Wang, T.~Kim, L.~Lu, and W.~Lee, ``Preventing use-after-free with dangling pointers nullification.'' in \emph{NDSS}, 2015.

\bibitem{dangsan}
E.~Van Der~Kouwe, V.~Nigade, and C.~Giuffrida, ``Dangsan: Scalable use-after-free detection,'' in \emph{Proceedings of the Twelfth European Conference on Computer Systems}, 2017.

\bibitem{markus}
S.~Ainsworth and T.~M. Jones, ``Markus: Drop-in use-after-free prevention for low-level languages,'' 2020.

\bibitem{dangzero}
F.~Gorter, K.~Koning, H.~Bos, and C.~Giuffrida, ``Dangzero: Efficient use-after-free detection via direct page table access,'' in \emph{Proceedings of the 2022 ACM SIGSAC Conference on Computer and Communications Security}, 2022.

\bibitem{ffmalloc}
\emph{Preventing Use-After-Free Attacks with Fast Forward Allocation.}, 2021.

\bibitem{cets}
S.~Nagarakatte, J.~Zhao, M.~M. Martin, and S.~Zdancewic, ``Cets: compiler enforced temporal safety for c,'' in \emph{Proceedings of the 2010 international symposium on Memory management}, 2010.

\bibitem{softbound}
------, ``Softbound: Highly compatible and complete spatial memory safety for c,'' in \emph{Proceedings of the 30th ACM SIGPLAN Conference on Programming Language Design and Implementation}, 2009.

\bibitem{in_fat_pointer}
S.~Xu, W.~Huang, and D.~Lie, ``In-fat pointer: hardware-assisted tagged-pointer spatial memory safety defense with subobject granularity protection,'' in \emph{Proceedings of the 26th ACM International Conference on Architectural Support for Programming Languages and Operating Systems}, 2021.

\bibitem{low_fat}
G.~J. Duck and R.~H. Yap, ``Heap bounds protection with low fat pointers,'' in \emph{Proceedings of the 25th International Conference on Compiler Construction}, 2016.

\bibitem{li2022pacmem}
Y.~Li, W.~Tan, Z.~Lv, S.~Yang, M.~Payer, Y.~Liu, and C.~Zhang, ``Pacmem: Enforcing spatial and temporal memory safety via arm pointer authentication,'' in \emph{Proceedings of the 2022 ACM SIGSAC Conference on Computer and Communications Security}, 2022.

\bibitem{kroes2018delta}
T.~Kroes, K.~Koning, E.~van~der Kouwe, H.~Bos, and C.~Giuffrida, ``Delta pointers: Buffer overflow checks without the checks,'' in \emph{Proceedings of the Thirteenth EuroSys Conference}, 2018, pp. 1--14.

\bibitem{dang2017oscar}
T.~H. Dang, P.~Maniatis, and D.~Wagner, ``Oscar: A practical $\{$Page-Permissions-Based$\}$ scheme for thwarting dangling pointers,'' in \emph{26th USENIX security symposium (USENIX security 17)}, 2017, pp. 815--832.

\bibitem{farkhani2021ptauth}
R.~M. Farkhani, M.~Ahmadi, and L.~Lu, ``$\{$PTAuth$\}$: Temporal memory safety via robust points-to authentication,'' in \emph{30th USENIX Security Symposium (USENIX Security 21)}, 2021, pp. 1037--1054.

\bibitem{saileshwar2022heapcheck}
G.~Saileshwar, R.~Boivie, T.~Chen, B.~Segal, and A.~Buyuktosunoglu, ``Heapcheck: Low-cost hardware support for memory safety,'' \emph{ACM Transactions on Architecture and Code Optimization (TACO)}, vol.~19, no.~1, pp. 1--24, 2022.

\bibitem{kim2020hardware}
Y.~Kim, J.~Lee, and H.~Kim, ``Hardware-based always-on heap memory safety,'' in \emph{2020 53rd Annual IEEE/ACM International Symposium on Microarchitecture (MICRO)}.\hskip 1em plus 0.5em minus 0.4em\relax IEEE, 2020, pp. 1153--1166.

\bibitem{woodruff2014cheri}
J.~Woodruff, R.~N. Watson, D.~Chisnall, S.~W. Moore, J.~Anderson, B.~Davis, B.~Laurie, P.~G. Neumann, R.~Norton, and M.~Roe, ``The cheri capability model: Revisiting risc in an age of risk,'' \emph{ACM SIGARCH Computer Architecture News}, vol.~42, no.~3, pp. 457--468, 2014.

\bibitem{zhang2019bogo}
T.~Zhang, D.~Lee, and C.~Jung, ``Bogo: Buy spatial memory safety, get temporal memory safety (almost) free,'' in \emph{Proceedings of the Twenty-Fourth International Conference on Architectural Support for Programming Languages and Operating Systems}, 2019, pp. 631--644.

\bibitem{asan_junxu}
Y.~Zhang, C.~Pang, G.~Portokalidis, N.~Triandopoulos, and J.~Xu, ``Debloating address sanitizer,'' in \emph{31st USENIX Security Symposium (USENIX Security 22)}, 2022.

\bibitem{zhang2021sanrazor}
J.~Zhang, S.~Wang, M.~Rigger, P.~He, and Z.~Su, ``Sanrazor: Reducing redundant sanitizer checks in c/c++ programs.'' in \emph{OSDI}, 2021.

\bibitem{camp_source}
``{The source code of CAMP},'' \url{https://github.com/cla7aye15I4nd/CAMP}.

\bibitem{magic_value}
J.~M. Mark~Dowd and J.~Schuh, ``{Magic Value: Potential Mitigations for Heap Overflow},'' \url{https://cwe.mitre.org/data/definitions/122.html}.

\bibitem{heap_cookie}
``{Heap Cookies for memory protection},'' \url{https://fuzzysecurity.com/tutorials/mr_me/3.html}.

\bibitem{redzone}
M.~Phillips, ``{Design of Redzone in Address Sanitizer},'' \url{https://github.com/google/sanitizers/wiki/AddressSanitizerAlgorithm}.

\bibitem{pointer_validation}
J.~M. Mark~Dowd and J.~Schuh, ``{Protect Out-Of-Bound by Validating Pointer},'' \url{https://cwe.mitre.org/data/definitions/823.html}.

\bibitem{delay_mem_free}
T.~Yamauchi and Y.~Ikegami, ``Heaprevolver: Delaying and randomizing timing of release of freed memory area to prevent use-after-free attacks,'' in \emph{Network and System Security: 10th International Conference, NSS 2016, Taipei, Taiwan, September 28-30, 2016, Proceedings 10}.\hskip 1em plus 0.5em minus 0.4em\relax Springer, 2016.

\bibitem{nullify_references}
O.~Corporation, ``{{Nullify references after reclaiming memory}},'' \url{https://docs.oracle.com/cd/E19159-01/819-3681/abebi/index.html}, 2010.

\bibitem{carat_native}
B.~Suchy, S.~Campanoni, N.~Hardavellas, and P.~Dinda, ``Carat: A case for virtual memory through compiler-and runtime-based address translation,'' in \emph{Proceedings of the 41st ACM SIGPLAN Conference on Programming Language Design and Implementation}, 2020.

\bibitem{carat_cake}
B.~Suchy, S.~Ghosh, D.~Kersnar, S.~Chai, Z.~Huang, A.~Nelson, M.~Cuevas, A.~Bernat, G.~Chaudhary, N.~Hardavellas \emph{et~al.}, ``Carat cake: Replacing paging via compiler/kernel cooperation,'' in \emph{Proceedings of the 27th ACM International Conference on Architectural Support for Programming Languages and Operating Systems}, 2022.

\bibitem{tcmalloc}
G.~Inc., ``{{Design of TCMalloc from Google}},'' \url{https://google.github.io/tcmalloc/overview.html}.

\bibitem{fixed_point}
V.~Berinde and F.~Takens, \emph{Iterative approximation of fixed points}.\hskip 1em plus 0.5em minus 0.4em\relax Springer, 2007, vol. 1912.

\bibitem{mruby_issue_3515}
C.~Smith, ``{{Heap use-after-free in mruby}},'' \url{https://github.com/mruby/mruby/issues/3515}.

\bibitem{chrome_issue_1325664}
``{{Issue 1325664: Security: pdfium use-after-free in v8}},'' \url{https://bugs.chromium.org/p/chromium/issues/detail?id=1325664}.

\bibitem{mruby_issue_5551}
``{{Heap-based Buffer Overflow in mruby}},'' \url{https://huntr.dev/bounties/4458e0b9-0ad3-4036-a032-1b3c4705b889/}.

\bibitem{cve_details}
``{{CVE details: The ultimate security vulnerability datasource}},'' \url{https://www.cvedetails.com/}.

\bibitem{cve_program}
``{{CVE program}},'' \url{https://cve.mitre.org/}.

\bibitem{vuln_db}
``{{VulnDB: The most comprehensive vulnerability database and timely source of intelligence available}},'' \url{https://vuldb.com/}.

\bibitem{cve-2021-26259}
``{CVE-2021-26259: A flaw was found in htmldoc in v1.9.12. Heap buffer overflow},'' \url{https://nvd.nist.gov/vuln/detail/CVE-2021-26259}.

\bibitem{duck2018effectivesan}
G.~J. Duck and R.~H. Yap, ``Effectivesan: type and memory error detection using dynamically typed c/c++,'' in \emph{Proceedings of the 39th ACM SIGPLAN Conference on Programming Language Design and Implementation}, 2018, pp. 181--195.

\bibitem{top_website}
``{Top Websites Ranking},'' \url{https://www.similarweb.com/top-websites/}.

\bibitem{prober}
H.~Liu, R.~Tian, B.~Ren, and T.~Liu, ``Prober: practically defending overflows with page protection,'' in \emph{Proceedings of the 35th IEEE/ACM International Conference on Automated Software Engineering}, 2020.

\bibitem{mimalloc}
Microsoft, ``{Daan Leijen. 2020. Mimalloc},'' \url{https://github.com/microsoft/mimalloc}.

\bibitem{Slimguard}
B.~Liu, P.~Olivier, and B.~Ravindran, ``Slimguard: A secure and memory-efficient heap allocator,'' in \emph{Proceedings of the 20th International Middleware Conference}, 2019.

\bibitem{scudo_allocator}
D.~V. Kostya~Serebryany, ``{Scudo Hardened Allocator},'' \url{https://llvm.org/docs/ScudoHardenedAllocator.html}.

\bibitem{undangle}
J.~Caballero, G.~Grieco, M.~Marron, and A.~Nappa, ``Undangle: early detection of dangling pointers in use-after-free and double-free vulnerabilities,'' in \emph{Proceedings of the 2012 International Symposium on Software Testing and Analysis}, 2012.

\bibitem{redfat}
G.~J. Duck, Y.~Zhang, and R.~H. Yap, ``Hardening binaries against more memory errors,'' in \emph{Proceedings of the Seventeenth European Conference on Computer Systems}, 2022, pp. 117--131.

\bibitem{jeon2020fuzzan}
Y.~Jeon, W.~Han, N.~Burow, and M.~Payer, ``$\{$FuZZan$\}$: Efficient sanitizer metadata design for fuzzing,'' in \emph{2020 USENIX Annual Technical Conference (USENIX ATC 20)}, 2020, pp. 249--263.

\bibitem{burow2018cup}
N.~Burow, D.~McKee, S.~A. Carr, and M.~Payer, ``Cup: Comprehensive user-space protection for c/c++,'' in \emph{Proceedings of the 2018 on Asia Conference on Computer and Communications Security}, 2018, pp. 381--392.

\bibitem{valgrind}
N.~Nethercote and J.~Seward, ``Valgrind: a framework for heavyweight dynamic binary instrumentation,'' \emph{ACM Sigplan notices}, vol.~42, no.~6, pp. 89--100, 2007.

\bibitem{sharifi2020chex86}
R.~Sharifi and A.~Venkat, ``Chex86: Context-sensitive enforcement of memory safety via microcode-enabled capabilities,'' in \emph{2020 ACM/IEEE 47th Annual International Symposium on Computer Architecture (ISCA)}.\hskip 1em plus 0.5em minus 0.4em\relax IEEE, 2020, pp. 762--775.

\end{thebibliography}
}

\appendix
% \newpage
% \section{Appendix}
% \subsection{Case study here}
% \vspace{-10pt}
\begin{table*}[]
\centering

\begin{tabular}{lcccccc}

\toprule
\multicolumn{1}{c}{\multirow{2}{*}{\textbf{Benchmarks}}} & \multicolumn{6}{c}{\textbf{Time}}                                                                                                                                             \\
\cline{2-7} \rule{0pt}{11pt}                            & \textbf{CAMP}                        & \textbf{ASAN$--$}               & \textbf{ASAN}                 & \textbf{ESAN}                 & \textbf{Softbound+CETS} & \textbf{MEMCHECK} \\
\midrule
400.perlbench  & 300.23\% & 256.68\% & 285.49\% & 629.16\% & -        & 3407.29\% \\
401.bzip2      & 48.90\%  & 43.98\%  & 48.22\%  & 114.70\% & 354.91\% & 912.38\%  \\
403.gcc        & 67.86\%  & 175.04\% & 173.97\% & 600.95\% & -        & 1877.21\% \\
429.mcf        & 27.25\%  & 24.06\%  & 34.85\%  & 158.27\% & 634.40\% & 303.66\%  \\
433.milc       & 9.04\%   & 38.82\%  & 48.64\%  & 50.76\%  & 239.02\% & 1194.67\% \\
445.gobmk      & 28.14\%  & 36.86\%  & 38.84\%  & 52.46\%  & 356.30\% & 2418.05\% \\
456.hmmer      & 119.45\% & 90.07\%  & 89.83\%  & 270.44\% & 477.35\% & 1647.21\% \\
458.sjeng      & 10.26\%  & 40.23\%  & 48.64\%  & 13.35\%  & 264.45\% & 2329.69\% \\
462.libquantum & 18.74\%  & 14.75\%  & 20.11\%  & 197.61\% & -        & 553.84\%  \\
464.h264ref    & 126.17\% & 89.88\%  & 125.67\% & 326.47\% & -        & 2689.47\% \\
470.lbm        & 7.39\%   & 25.82\%  & 28.41\%  & 51.63\%  & 141.17\% & 5236.58\% \\
482.sphinx3    & 114.08\% & 44.70\%  & 52.97\%  & 199.55\% & -        & 4207.37\% \\
444.namd       & 90.43\%  & 75.60\%  & 82.01\%  & 57.64\%  & -        & 4076.65\% \\
450.soplex     & 64.42\%  & 42.82\%  & 44.45\%  & 128.66\% & -        & 1518.52\% \\
453.povray     & 113.95\% & 105.89\% & 150.95\% & 266.29\% & -        & 5385.34\% \\
473.astar      & 112.80\% & 30.62\%  & 37.97\%  & 80.75\%  & -        & 1085.92\% \\
483.xalancbmk  & 297.90\% & 166.85\% & 203.05\% & 48.80\%  & -        & 5158.20\% \\
\midrule
Geomean  & 54.92\% & 56.77\% & 67.02\% & 123.08\% & 319.75\% & 1990.02\%
\\
\bottomrule

\end{tabular}

\caption{Time overhead of CAMP, ASAN\code{--}, ASAN, ESAN, Softbound+CETS, MemCheck on the SPEC CPU2006. "-" means the case where the tool failed to run the benchmark.}
\label{tab:time-overhead-cpu2006}
\end{table*}

\begin{table*}[t]
\centering

\begin{tabular}{lcccccc}

\toprule
\multicolumn{1}{c}{\multirow{2}{*}{\textbf{Benchmarks}}} & \multicolumn{6}{c}{\textbf{Memory}}                                                                                                                                             \\
\cline{2-7} \rule{0pt}{11pt}                            & \textbf{CAMP}                        & \textbf{ASAN$--$}               & \textbf{ASAN}                 & \textbf{ESAN}                 & \textbf{Softbound+CETS} & \textbf{MEMCHECK} \\
\midrule
400.perlbench  & 1522.09\% & 339.64\%  & 285.49\% & 1.90\%   &  -       & 163.14\%  \\
401.bzip2      & 0.05\%    & 2.75\%    & 48.22\%  & -0.98\%  & 127.90\% & 33.59\%   \\
403.gcc        & 109.95\%  & 187.98\%  & 173.97\% & -6.78\%  & -        & 44.12\%   \\
429.mcf        & 51.66\%   & 12.02\%   & 34.85\%  & -0.56\%  & 396.77\% & 2.29\%    \\
433.milc       & 379.00\%  & 36.77\%   & 48.64\%  & -1.58\%  & 89.01\%  & 10.50\%   \\
445.gobmk      & 143.05\%  & 612.69\%  & 38.84\%  & -9.29\%  & 638.64\% & 230.80\%  \\
456.hmmer      & 4.32\%    & 1061.04\% & 89.83\%  & -39.34\% & -12.09\% & 197.22\%  \\
458.sjeng      & -0.10\%   & -1.69\%   & 48.64\%  & -3.26\%  & 1.18\%   & 44.82\%   \\
462.libquantum & -0.41\%   & 185.86\%  & 20.11\%  & -21.56\% &    -     & 48.34\%   \\
464.h264ref    & 18.82\%   & 330.51\%  & 125.67\% & -19.84\% &     -    & 120.71\%  \\
470.lbm        & -0.01\%   & 11.47\%   & 28.41\%  & -1.41\%  & -1.55\%  & 34.39\%   \\
482.sphinx3    & 4093.04\% & 650.28\%  & 52.97\%  & -4.03\%  &   -      & 240.28\%  \\
444.namd       & 3.63\%    & 9.93\%    & 82.01\%  & -8.45\%  &     -     & 101.86\%  \\
450.soplex     & 12.65\%   & 44.26\%   & 44.45\%  & -20.79\% &    -      & 21.20\%   \\
453.povray     & 7101.36\% & 2020.28\% & 150.95\% & -12.61\% &     -     & 1188.33\% \\
473.astar      & 1840.76\% & 176.09\%  & 37.97\%  & 5.29\%   &     -     & 50.27\%   \\
483.xalancbmk  & 1800.53\% & 187.89\%  & 203.05\% & 14.54\%  &   -       & 77.51\%   \\
\midrule
Geomean & 237.67\% & 181.81\% & 180.94\% & -8.45\% & 102.25\% & 97.14\%
\\
\bottomrule

\end{tabular}

\caption{Memory overhead of CAMP, ASAN\code{--}, ASAN, ESAN, Softbound+CETS, MemCheck on the SPEC CPU2006. "-" means the case where the tool failed to run the benchmark.}
\label{tab:memory-overhead-cpu2006}
\end{table*}
\end{document}